%% file: arxiv.tex
\documentclass[reqno,12pt]{amsart}

\usepackage[a4paper, 
            left=1.in,
            right=1in,
            top=1in,
            bottom=1in, 
            footskip=.25in]{geometry}

\usepackage[utf8]{inputenc}
\usepackage{microtype}
\usepackage{graphicx}
\usepackage{amssymb} 
\usepackage{hyperref}
 
\usepackage{thmtools} 
\usepackage{thm-restate}

\usepackage{microtype}

\theoremstyle{plain}
\newtheorem{thm}{}[section]
\newtheorem{lemma}[thm]{Lemma}
\newtheorem{proposition}[thm]{Proposition} 

\newtheorem{theorem}[thm]{Theorem}
\newtheorem{corollary}[thm]{Corollary}
\theoremstyle{remark}

 \newtheorem{remark}[thm]{Remark} 
 \newtheorem{example}[thm]{Example}

\usepackage{centernot}  
\usepackage{xspace}  

\usepackage{hyperref}
 
\usepackage[scr=boondoxo]{mathalpha}  
\newcommand{\coloneqq}{\mathrel{\mathop:}\mathrel{\mkern-1.2mu}=} 
\newcommand{\struct}[1]{\mathfrak{#1}}    
\newcommand{\CSP}{\ensuremath{\mathrm{CSP}}\xspace}   
  
\newcommand{\Age}{\ensuremath{\mathrm{age}}\xspace}   
\newcommand{\fm}{\ensuremath{\mathrm{fm}}\xspace}      
    
\newcommand{\NP}{{\textup{\textsf{NP}}}}    
  
\newcommand{\NEXPTIME}{{\textup{\textsf{NEXPTIME}}}}   
\newcommand{\coNEXPTIME}{{\textup{\textsf{coNEXPTIME}}}} 
\newcommand{\TWONEXPTIME}{{\textup{\textsf{2NEXPTIME}}}}      
\newcommand{\Forb}{\ensuremath{\mathrm{Forb}}\xspace} 
\newcommand{\fancyfont}[1]{\mathscr{#1}}   
\newcommand{\ncolors}{\ensuremath{n\textup{-colours}}\xspace}

\begin{document}

\title{Finitely Bounded Homogeneity Turned Inside-Out} 
 
\author{Jakub Rydval}
\address{Institut f\"{u}r Diskrete Mathematik und Geometrie, FG Algebra, TU Wien, Austria}
\email{jakub.rydval@tuwien.ac.at}

 \begin{abstract}
Deciding the amalgamation property for a given class of finite structures is an important subroutine in classifying countable finitely homogeneous structures.
We study the computational complexity of the amalgamation decision problem for finitely bounded classes, i.e., classes specified by a finite set of forbidden finite substructures, or equivalently by a finite set of universal axioms. 

We link the amalgamation decision problem to the problem of testing the containment between the reducts of two given finitely bounded amalgamation classes to a given common subset of their signatures. 
On the one hand, this link enables polynomial-time reductions from various decision problems that can be represented within the reduct containment problem for finitely bounded amalgamation classes, e.g., the 2-exponential square tiling problem, leading to a new lower bound for the complexity of the amalgamation decision problem: $\TWONEXPTIME$-hardness.
On the other hand, the link also allows us to show that the amalgamation decision problem is decidable under the assumption that every finitely bounded strong amalgamation class has a computable finitely bounded Ramsey expansion.
The runtime of our conditional decision procedure depends 2-exponentially on the size of a minimal Ramsey expansion.
We subsequently prove that the closely related problem of testing homogenizability is already undecidable, by a polynomial-time reduction from the regularity of context-free languages.

Our results indicate that the relationship between finitely bounded amalgamation classes and arbitrary finitely bounded classes shares similarities with the relationship between regular grammars and context-free grammars. 
A key difference is that the regularity of context-free grammars can be tested in linear time, while the problem of testing the amalgamation property for finitely bounded classes is \TWONEXPTIME-hard. 
\end{abstract}  

\thanks{This research was funded in whole or in part by the Austrian Science Fund (FWF) [I 5948]. For the purpose of Open Access, the authors have applied a CC BY public copyright licence to any Author Accepted Manuscript (AAM) version arising from this submission.
\newline A conference version of this material appeared in the Proceedings of the 51st EATCS International Colloquium on Automata, Languages and Programming (ICALP), \#150, 1-20, 2024~\cite{rydval2024homogeneity}.}

\maketitle
  
 \input{content}

\section*{Acknowledgements} The author thanks Manuel Bodirsky, Simon Kn\"auer, Jakub Opr\v{s}al, Paolo Marimon, Michael Pinsker, Mat\v{e}j Kone\v{c}n\'{y}, Alexey Barsukov, and Moritz Sch\"{o}bi for inspiring discussions on the topic, and the anonymous reviewers for many helpful suggestions (in particular the one who provided six pages worth of valuable comments).

 \appendix

\input{appendix}

\bibliographystyle{abbrv}
\bibliography{homogeneity_and_homogenizability}

\end{document}

%% file: content.tex
\section{Introduction}
A relational structure $\struct{B}$ is called \emph{homogeneous} if every isomorphism between two of its finite substructures extends to an automorphism of the entire structure $\struct{B}$; we say that $\struct{B}$ is \emph{finitely homogeneous} if it is homogeneous and its signature is finite.
In the case where $\struct{B}$ is additionally countable, it is uniquely determined up to isomorphism by its \emph{age}, i.e., the class of all finite structures embeddable into $\struct{B}$~\cite{Hodges}.
The classification problem for countable finitely homogeneous structures is notoriously difficult~\cite{Latka_1994,AkhtarLachlan95,knight2006shrinking,cherlin2022homogeneous}, with only a handful of published classification results; see, e.g., Cherlin's classification of countable homogeneous directed graphs~\cite{cherlin1998classification,cherlin1993homogeneous} stretching over almost 160 pages.
It is moreover known that there are uncountably many different countable homogeneous structures already among directed graphs; this result is due to Henson~\cite{henson1972countable}.
Hence, it is not even possible to check concrete instances of the classification problem on a computer, because most finitely homogeneous structures do not have a finite presentation.
On the other hand, taking the algorithmic perspective does make sense under additional structural restrictions that do not allow for constructions such as the one of Henson~\cite{henson1972countable} to take place.
One such restriction, which is central to the present article, is called finite boundedness. 
 
We say that a class $\mathcal{K}$ of finite structures over a finite relational signature $\tau$ is \emph{finitely bounded}~\cite{MacphersonSurvey} if there exists a finite set $\mathcal{B}$ of finite $\tau$-structures (\emph{bounds}) such that $\mathcal{K}$ consists of all finite $\tau$-structures which do not embed any member of $\mathcal{B}$; we then write $\mathcal{K}=\Forb_{e}(\mathcal{B})$. 
A countable $\tau$-structure is \emph{finitely bounded} if its age has this property.  
By Fra\"iss\'e's theorem, a finitely bounded class $\mathcal{K}$ forms the age of an (up to isomorphism unique) countable finitely bounded homogeneous structure if and only if $\mathcal{K}$ has the \emph{Amalgamation Property} (AP), which states that each two members $\struct{B}_1,\struct{B}_2\in \mathcal{K}$ can be embedded into a larger structure $\struct{C} \in \mathcal{K}$ while preserving the intersection of $\struct{B}_1$ and $\struct{B}_2$.
We allow the amalgamation over empty structures, i.e., where the domains of $\struct{B}_1$ and $\struct{B}_2$ do not intersect.
We further distinguish the \emph{Strong Amalgamation Property} (SAP) where $\struct{C}$ can always be obtained by adding tuples to the relations of the union of $\struct{B}_1$ and $\struct{B}_2$, and the \emph{Free Amalgamation Property} (FAP) where $\struct{C}$ can always be chosen as the union of $\struct{B}_1$ and $\struct{B}_2$. 
A thorough definition of these properties can be found in Section~\ref{section:fraisse}. 
The fact that the classification problem for countable finitely bounded homogeneous structures only depends on whether a given set of bounds specifies a class of structures with the AP implies that it can be expressed as a computational decision problem:   
 
\medskip   
\noindent  \underline{\textbf{Amalgamation Decision Problem (ADP)}}  \hfill // Input type: sets of bounds  \\
 \setlength{\tabcolsep}{0pt}    \begin{tabular}{ll}  
  INSTANCE: & \,  A finite set $\mathcal{B}$ of finite structures over a finite relational signature $\tau$. \\ 
      QUESTION: & \,  Does $\Forb_{e}(\mathcal{B})$ have the AP?  
\end{tabular}   
\medskip

In the present article, we initiate a systematic study of the computational complexity of the ADP. 
To enable a more fine-grained complexity-theoretic approach, we  adopt a different but equivalent definition of finite boundedness, namely, in terms of definability by a universal first-order sentence. 
For a universal first-order sentence $\Phi$, we denote the class of its finite models by $\fm(\Phi)$.
The ADP can now be restated as follows:

\medskip    
\noindent \underline{\textbf{Amalgamation Decision Problem (ADP)}} \hfill // Input type: universal axioms  
 \setlength{\tabcolsep}{0pt}    \begin{tabular}{ll}  
  INSTANCE: & \,  A universal first-order sentence $\Phi$ over a finite relational signature $\tau$. \\ 
      QUESTION: & \,   Does $\fm(\Phi)$ have the AP? 
\end{tabular}   
 \medskip   

\noindent The advantage is that now the inputs to the ADP are better behaved under basic modifications, e.g., their size does not increase exponentially if we add a fresh symbol to the signature.
Here, by the \emph{size of a set of bounds} $\mathcal{B}$ we mean the sum of the sizes of all structures in $\mathcal{B}$, where the size of a structure is the sum of the cardinalities of the domain and all relations; by the \emph{size of a universal first-order sentence} $\Phi$ we mean the number of symbols in $\Phi$ and its signature.
The downside is that the equivalence of the two definitions only holds up to a single-exponential blow-up in one direction, 
we elaborate on this in the paragraph below.

Given a finite set of bounds $\mathcal{B}$, we can obtain a universal first-order sentence $\Phi$ of size polynomial in the size of $\mathcal{B}$ satisfying $\fm(\Phi) = \Forb_{e}(\mathcal{B})$  simply by describing each structure in $\mathcal{B}$ up to isomorphism using a quantifier-free formula.
However, given a universal sentence $\Phi$, it can be that a smallest $\mathcal{B}$ satisfying $\fm(\Phi) = \Forb_{e}(\mathcal{B})$ is of size single-exponential in the size of $\Phi$. 
The reason is that obtaining $\mathcal{B}$ from $\Phi$ is comparable to rewriting the quantifier-free part of $\Phi$ in CNF.
For example, the class of all directed graphs without any directed paths of length $n$ is definable by 
$
\Phi\coloneqq \forall x_1,\dots, x_{n+1}  \ \neg  ( \bigwedge\nolimits_{i=1}^n E(x_i,x_{i+1}) \wedge \bigwedge\nolimits_{i\neq j} x_i\neq x_j  ).
$
However, every set of bounds for this class must necessarily contain up to isomorphism all directed graphs on $n+1$ vertices containing a directed path of length $n$.
The number of such graphs is exponential in $n$.
Nevertheless, the size difference between the two input types essentially only constitutes a single-exponential offset in the complexity and therefore does not have a significant impact on our results.

\subsection{Lower bounds}
We show that the ADP admits an efficient reduction from the problem of testing the containment between the reducts of two given finitely bounded amalgamation classes to a given common subset of their signatures.
Note that, using a standard compactness argument (Lemma~\ref{lemma:compactness}), the latter is equivalent to testing the embeddability between reducts of finitely bounded homogeneous structures. 
In fact, we give a reduction from more general reduct containment problem where only one of the two finitely bounded classes in the input needs to be an amalgamation class.

For a class $\mathcal{K}$ of structures with a common signature $\sigma$ and a subset $\tau\subseteq \sigma$, we write $\mathcal{K}|^{\tau}$ for the class of the $\tau$-reducts of all structures in $\mathcal{K}$.

\medskip   
\noindent  \underline{\textbf{Reduct Containment Problem (RCP)}}  \\
  \setlength{\tabcolsep}{0pt}    \begin{tabular}{ll}  
  INSTANCE: \, &    Universal first-order sentences $\Phi_1$ and $\Phi_2$ over finite relational \\
     \, &    signatures $\sigma_1$ and $\sigma_2$ and a common subset $\tau\subseteq \sigma_1\cap \sigma_2$. \\
      QUESTION: \, &     Does $\fm(\Phi_1)|^{\tau}\subseteq \fm(\Phi_2)|^{\tau}$ hold?  
\end{tabular}   
\medskip

We distinguish the following restrictions of the RCP, which involve varying degrees of promise of the amalgamation property for the input.
For each instance $(\Phi_1,\Phi_2,\tau)$ of the RCP, we are promised that:

\medskip   
\noindent
\underline{\textbf{(freely/strongly) semi-amalgamated RCP}} \\
 \setlength{\tabcolsep}{0pt}    \begin{tabular}{ll}  
  PROMISE: \, &  $\fm(\Phi_1)$ has the (free/strong) amalgamation property;
\end{tabular}  
\medskip 

\noindent\underline{\textbf{(freely/strongly)  amalgamated RCP}} \\
 \setlength{\tabcolsep}{0pt}    \begin{tabular}{ll}  
  PROMISE: \, &  $\fm(\Phi_1)$ and $\fm(\Phi_2)$ have the (free/strong) amalgamation property.
\end{tabular} 
\medskip  

A proof of the following theorem can be found at the beginning of Section~\ref{section:inside_out}.
 
\begin{restatable}{theorem}{firstinsideout}  \label{thm:inside_out}    There exists a polynomial-time computable function $\Omega$ mapping each pair $(\Phi_1,\Phi_2)$ of universal first-order sentences and each common subset $\tau$ of their signatures to a universal sentence $\Phi\coloneqq \Omega(\Phi_1,\Phi_2,\tau)$ such that the following are equivalent:
\begin{enumerate}  
    \item  \label{item:inside_out2} $\fm(\Phi)$ has the SAP;
    \item \label{item:inside_out3} $\fm(\Phi)$ has the AP; 
    \item \label{item:inside_out1}  $\fm(\Phi_1)$ has the SAP and $\fm(\Phi_1)|^{\tau}\subseteq \fm(\Phi_2)|^{\tau}$.
\end{enumerate} 
\end{restatable}  
As a consequence of Theorem~\ref{thm:inside_out}, there exists a polynomial-time reduction from the strongly semi-amalgamated RCP to the ADP.
There is a natural way to extend this reduction to the semi-amalgamated RCP --- using the following theorem.
A proof of Theorem~\ref{thm:SAP_lemma} below can be found in Section~\ref{section:inside_out}.
\begin{restatable}{theorem}{saplemma}    \label{thm:SAP_lemma} 
   There exists a polynomial-time computable function $\Gamma$ mapping each universal first-order sentence $\Phi$ to a universal first-order sentence $\Gamma(\Phi)$ over the signature of $\Phi$ expanded by a fresh binary symbol $E$ such that 
 \begin{enumerate}
     \item \label{item:sap1} $\fm(\Phi)$ has the AP if and only if $\fm(\Gamma(\Phi))$ has the SAP. 
 \end{enumerate} 
   Moreover, if $\Phi_1$ and $\Phi_2$ are universal first-order sentences, $\tau$ is a common subset of their signatures, and $E$ is the fresh binary symbol coming from $\Gamma$, then  
\begin{enumerate}
  \setcounter{enumi}{1}
  \item \label{item:sap2} $\fm(\Phi_1)|^{\tau}\subseteq \fm(\Phi_2)|^{\tau}$ if and only if $\fm(\Gamma(\Phi_1))|^{\tau\cup \{E\}}\subseteq \fm(\Gamma(\Phi_2))|^{\tau\cup \{E\}}.$
\end{enumerate} 
\end{restatable}     
Theorem~\ref{thm:inside_out} and Theorem~\ref{thm:SAP_lemma} together imply that, in order to understand the complexity of the ADP, the problem of recognizing amalgamation classes among arbitrary finitely bounded classes, we must first understand the complexity of the semi-amalgamated RCP, the problem of testing the containment between  reducts of finitely bounded classes $\mathcal{K}_1$ and $\mathcal{K}_2$, where $\mathcal{K}_1$ is promised to be an amalgamation class.
Now, \emph{without} the promise of the AP for $\mathcal{K}_1$, the  reduct containment is in general undecidable, since it admits a polynomial-time reduction from the undecidable~\cite[Theorem~8.4.4]{harrison1978introduction} containment problem for context-free languages; see Section~\ref{section:encoding} for a suitable encoding into finitely bounded classes.
Therefore, if the ADP is decidable, then the promise of the AP for $\mathcal{K}_1$ must offer a significant advantage in the analysis of $\mathcal{K}_1|^\tau \subseteq \mathcal{K}_2|^\tau$.
However, in order to be able to use the advantage provided by the promise, we would first have to understand the promise, which is the entire point of the ADP.
The circular nature of this phenomenon motivates the name ``Finitely Bounded Homogeneity Turned Inside-Out''. 

The next corollary is a direct consequence of Theorem~\ref{thm:inside_out} and Theorem~\ref{thm:SAP_lemma}.
\begin{corollary}  \label{cor:summary} The following decision problems are polynomial-time equivalent:
\begin{itemize} 
    \item  Given a universal first-order sentence $\Phi$ over a finite relational signature, decide whether $\fm(\Phi)$ has the (S)AP;  
    \item  Given universal first-order sentences $\Phi_1$ and $\Phi_2$ over finite relational signatures $\sigma_1$ and $\sigma_2$, respectively, and a subset $\tau\subseteq \sigma_1\cap \sigma_2$, decide whether  
\begin{center}
      $\fm(\Phi_1)$  has the (S)AP \quad and \quad $\fm(\Phi_1)|^{\tau} \subseteq \fm(\Phi_2)|^{\tau}$.
\end{center} 
\end{itemize}   
\end{corollary} 

 Theorem~\ref{thm:inside_out} would become a trivial statement if it turned out that the promise of the SAP trivializes the RCP.
This is not the case, as we will see in the next theorem; in fact, we propose Theorem~\ref{thm:inside_out} as a robust tool for obtaining lower bounds for the ADP.
A proof of Theorem~\ref{thm:containment_hard} below can be found in Section~\ref{section:cont_hard}.

\begin{restatable}{theorem}{containmenthard}   \label{thm:containment_hard} The freely amalgamated RCP is \TWONEXPTIME-hard.
\end{restatable}   
\begin{corollary} \label{cor:hardness} The ADP (universal axioms) is \TWONEXPTIME-hard.
\end{corollary}   
We remark that Corollary~\ref{cor:hardness} has a counterpart in the setting where instances are specified by sets of bounds; here, the lower bound drops to $\NEXPTIME$ as a result of the input conversion (see Remark~\ref{rk:bounds}).
\begin{restatable}{corollary}{hardnesssetsofbounds}   \label{cor:input_bounds} The ADP (sets of bounds) is \NEXPTIME-hard.
\end{restatable}   
Finally, we comment on the fact that Theorem~\ref{thm:inside_out} does not impose any restrictions on $\fm(\Phi_2)$, which indicates that the lower bound stemming from Theorem~\ref{thm:containment_hard} might be suboptimal.
There is an abundance of instances of the strongly semi-amalgamated RCP which cannot be represented by any instance of the (strongly) amalgamated RCP; one example is given below.
However, such instances do not seem to encode any computationally harder problems.  

\begin{example} Consider the following two sentences $\Phi_1$ and $\Phi_2$ over the binary signatures $\sigma_1=\{E,O\}$ and $\sigma_2= \{E,P,O,K\}$, respectively: 
\begin{align*}
    \Phi_1 \coloneqq  \forall x,y,z \, &\big( E(x,y) \Rightarrow O(x,y)   \big) \wedge \text{``}O\text{ is irreflexive and transitive''}      
\\
  \Phi_2 \coloneqq   \forall x,y,z \, & \big( E(x,y) \Rightarrow P(x,y) \big) \wedge \big( P(x,y) \Rightarrow O(x,y) \big)   \\
   {} \wedge \, &  \big(P(x, y) \wedge P(x, z) \Rightarrow K(y, z) \big)  \wedge \big(K(x, y) \wedge P(x, z)  \Rightarrow  P(y, z)\big)\\ 
  {} \wedge \, &  \text{``} O \text{ is irreflexive and transitive and } K \text{ is an equivalence relation''}      
\end{align*} 
Let $\tau\coloneqq \{E\}$.
Then $\fm(\Phi_1)|^{\tau}$ defines the class of all finite directed graphs that homomorphically map to a finite strict partial order, and $\fm(\Phi_2)|^{\tau}$ defines the class of all finite directed graphs that homomorphically map to a finite directed path~\cite[Sec. 5.8]{Book}.
It can be easily seen that $\fm(\Phi_1)|^{\tau}\subseteq \fm(\Phi_2)|^{\tau}$ is  false.
It is also easy to verify that $\fm(\Phi_1)$ has the SAP (finite strict partial orders form a strong amalgamation class), but there is  no universal sentence $\Phi'_2$ with $\fm(\Phi'_2)|^{\tau}=\fm(\Phi_2)|^{\tau}$ and such that 
$\fm(\Phi'_2)$ would have the AP~\cite[Sec. 5.8]{Book}.
We say that the class $\fm(\Phi_2)|^{\tau}$ is not homogenizable; we discuss homogenizability in~Section~\ref{section:homogenizability}.
\end{example}

\subsection{Towards decidability} 

The question of decidability of the ADP has been considered many times in the context of the Lachlan-Cherlin programme for countable homogeneous structures, and is known to have a positive answer in the case of binary signatures~\cite{lachlan1986homogeneous,bodirsky2020asnp}.
An inspection of the decidability result for binary signatures reveals that the ADP over binary signatures is decidable in \coNEXPTIME\  (Proposition~\ref{cor:conexptime}).
Hence, our \TWONEXPTIME\ lower bound (Corollary~\ref{cor:hardness}) shows a relative increase in the complexity compared to the binary case.
The idea behind the algorithm in the binary case is rather simple, so we do not expect the \coNEXPTIME\ upper bound to be sharp; the best lower bound we currently have is $\Pi^p_3$-hardness (Proposition~\ref{prop:binary_intro}).  
We remark that the \coNEXPTIME\ upper bound drops to $\Pi^p_2$ if the 
inputs are specified by sets of bounds instead of universal first-order sentences (Theorem~15 in~\cite{baader2022using}).

\medskip 
\noindent\textbf{Question 1:} What is the precise complexity of the ADP over binary signatures?
\medskip

We do not provide any concrete upper bounds on the complexity of the ADP for arbitrary finite signatures, but we connect it to a different but related problem stemming from questions in structural Ramsey theory.
To this end, we first give an auxiliary result --- the second part of the ``inside-out'' correspondence --- \emph{almost} showing that the ADP admits an efficient reduction back to the strongly amalgamated RCP.
A proof of the following theorem can be found at the end of Section~\ref{section:inside_out}. 

\begin{restatable}{theorem}{secondinsideout} 
 \label{thm:inside_out_2}   There exists a polynomial-time computable function $\Delta$ mapping each universal first-order sentence $\Phi$ over a finite relational signature $\rho$ to a pair of universal first-order sentences $(\Phi_1, \Phi_2)\coloneqq \Delta(\Phi)$ over finite relational signatures $\tau$ and $\sigma$, respectively, with $\rho \subseteq \tau \subseteq \sigma$ and such that the following are equivalent:
\begin{enumerate}  
    \item \label{item:in_out_1} $\fm(\Phi)$ has the SAP;
    \item \label{item:in_out_2} $\fm(\Phi_1)$ has the SAP;  
    \item \label{item:in_out_3} $\fm(\Phi_2)$ has the SAP; 
    \item \label{item:in_out_4} $\fm(\Phi_1) \subseteq \fm(\Phi_2)|^{\tau}$.
\end{enumerate} 
\end{restatable}

Now consider the following \emph{faulty} reduction from the ADP to the strongly amalgamated RCP: on an input $\Phi$ to the ADP, we
first apply $\Gamma$ from Theorem~\ref{thm:SAP_lemma}, then $\Delta$ from Theorem~\ref{thm:inside_out_2}, and finally forward the result to an oracle for the strongly amalgamated RCP.
The obvious issue here is the possible occurrence of false positives; an oracle for the strongly amalgamated RCP might return \emph{yes} on inputs where $\fm(\Phi_1) \subseteq \fm(\Phi_2)|^{\tau}$ does not hold but $\fm(\Phi_1)$ and $\fm(\Phi_2)$ do not have the SAP.
We leave the following question open.

\medskip 
\noindent\textbf{Question 2:} Is the ADP over arbitrary finite signatures reducible in polynomial time to the strongly (semi-)amalgamated RCP?
\medskip 

The issue of false positives can be circumvented under the assumption that every finitely bounded strong amalgamation class has a computable finitely bounded \emph{Ramsey expansion}, which motivates the formulation of Theorem~\ref{theorem:ramsey}.
By a Ramsey expansion of an amalgamation class $\mathcal{K}$ of $\tau$-structures, we mean a class $\mathcal{K}^+$ with $\mathcal{K}^+|^{\tau}=\mathcal{K}$ and which has the amalgamation and the \emph{Ramsey property} (RP).
The definition of the RP can be found in Section~\ref{section:prelims_logic}. 
Reformulated using Fra\"{i}ss\'{e}'s theorem, we want an expansion of our Fra\"{i}ss\'{e} limit which is again a Fra\"{i}ss\'{e} limit and whose age has the RP.
The question whether such an expansion always exists was left open in~\cite{bodirsky_pinsker_tsankov} in a considerably more general setting than in the present article but without the requirement of computability; see also Conjecture~1 in \cite{the2014survey}.
While the most general version of this question was answered negatively in~\cite{evans2019automorphism}, it remains open for arbitrary amalgamation classes over finite relational signatures (see Question 7.1 in~\cite{evans2019automorphism}).
For finitely bounded amalgamation classes, the question was formulated as a conjecture in Bodirsky's book on infinite-domain constraint satisfaction~\cite{Book}.  
A proof of Theorem~\ref{theorem:ramsey} below can be found in Section~\ref{section:ramsey}.

\begin{restatable}{theorem}{ramsey} 
\label{theorem:ramsey}
Suppose that there exists a function $g\colon \mathbb{N} \rightarrow \mathbb{N}$ such that every finitely bounded strong amalgamation class 
specified by a universal sentence of size $n$ has a finitely bounded Ramsey expansion specified by a universal sentence of size $g(n)$. 
Then the ADP can be solved non-deterministically in time $\mathcal{O}\bigl(2^{2^{g( p(n))}}\bigr)$, where $p$ is some fixed polynomial. 
\end{restatable}  

 \medskip 
\noindent\textbf{Question 3:} Is there a computable function $g\colon \mathbb{N} \rightarrow \mathbb{N}$ such that every finitely bounded strong amalgamation class 
specified by a universal sentence of size $n$ has a finitely bounded Ramsey expansion specified by a universal sentence of size $g(n)$?
\medskip 

   If the ADP turns out to be undecidable, then, by Theorem~\ref{theorem:ramsey}, the answer to the above question is negative (turning an undecidability proof into a negative Ramsey-theoretic result).
However, purely on the basis of empirical evidence for the complexity of the  containment problem in various settings linked to the ADP~\cite{bodirsky2021proof,bourhis2016containment,barsukov2025containment} and the typical sizes of finitely bounded Ramsey expansions~\cite{hubivcka2019all}, we would rather like to interpret Theorem~\ref{theorem:ramsey} the other way around and speculate that the ADP is decidable.  
We contrast this optimistic outlook with the fact that there are concrete examples of finitely bounded strong amalgamation classes for which no Ramsey expansion is known; see~\cite[Sec.~8.1.1]{hubivcka2025twenty}.
%

\medskip 
\noindent\textbf{Question 4:} Is the ADP  in \TWONEXPTIME?

\subsection{Homogenizability} \label{section:homogenizability}
Relaxing the question in the ADP to whether a given class is a reduct of a finitely bounded amalgamation class yields a fundamentally different problem, related to the question of homogenizability.  
The natural set of instances for this problem are the sentences of the logic \emph{Strict NP} (SNP)~\cite{kolaitis1987decision, papadimitriou1988optimization, feder1998computational}.
SNP is obtained from the universal fragment of first-order logic over relational signatures by allowing existential quantification over relation symbols at the beginning of the quantifier prefix; in particular, universal first-order formulas are SNP formulas.
As before, for an SNP sentence $\Phi$, we denote the class of all its finite models by $\fm(\Phi)$.

A countable structure $\struct{B}$ with a finite relational signature is \emph{homogenizable}~\cite{ahlman2016homogenizable,atserias2016non,Covington,hubivcka2019all} if it has a homogeneous expansion $\struct{B}^+$ by finitely many first-order definable relations.  
We say that $\struct{B}$ is \emph{finitely bounded-homogenizable} if $\struct{B}^+$ can 
be chosen finitely bounded. 
Sufficient conditions for finitely bounded-homogenizability were obtained by Hubi\v{c}ka and Ne\v{s}et\v{r}il~\cite{Hubicka-Nesetril}, generalizing previous work of Cherlin, Shelah, and Shi~\cite{cherlin1999universal}. 
We say that $\struct{B}$ is \emph{weakly} (\emph{finitely bounded-}) \emph{homogenizable} if it has any (finitely bounded) homogeneous expansion $\struct{B}^+$, i.e., where we do not impose any restrictions on how the relations of $\struct{B}^+$ are defined. 
In analogy to finitely bounded homogeneity, the computational decision problem associated with testing weak finitely bounded-homogenizability can be formulated using the AP:

\medskip    
\noindent   \underline{\textbf{Weak Homogenizability Problem (WHP)}} \\
\setlength{\tabcolsep}{0pt}    \begin{tabular}{ll}  
  INSTANCE: & \,  An SNP sentence $\Phi$ over a finite relational signature $\tau$. \\ 
      QUESTION: & \,  Does there exist a universal first-order sentence $\Phi^+$ such that 
\end{tabular}  
 \begin{center} 
$\fm(\Phi)=\fm(\Phi^{+})|^{\tau}$ \quad and \quad  $\fm(\Phi^{+})$ has the AP?
\end{center}     
\medskip 

We show that the WHP is undecidable, even when the instances are restriced to the Datalog fragment of SNP and use at most binary relation symbols~(Theorem~\ref{thm:undecidability}).  
Our proof also applies to the standard version of (finitely bounded-) homogenizability, where the homogeneous expansion must be first-order definable in its $\tau$-reduct.
As a byproduct of our proof, we also get the undecidability of some other properties for Datalog sentences, e.g., the questions whether a given Datalog sentence can be rewritten in monadic Datalog,  whether it defines a structure with a homogeneous Ramsey expansion in a finite relational signature, or whether it solves some finite-domain \emph{Constraint Satisfaction Problem} (CSP). 
We believe that these will be of independent interest to computer scientists. 

A CSP is the membership problem for the class of all finite structures which homomorphically map to some fixed structure $\struct{B}$ over a finite relational signature; this class is denoted by $\CSP(\struct{B})$. 
Note that a CSP can be parametrized by many different structures $\struct{B}$; we call a CSP~\emph{finite-domain} if $\struct{B}$ can be chosen finite  and \emph{infinite-domain} otherwise.

To keep our undecidability result (Theorem~\ref{thm:undecidability}) as general as possible, we formulate it as a statement about a promise relaxation of the various questions from above, i.e., where a subclass and a superclass of the positive instances are being separated from each other with the promise that the input never belongs to the complement of the subclass within the superclass.
More specifically, we formulate our main undecidability result as a statement of the form ``the question whether X or not even Y is undecidable.'' 
This is a compact way for writing that both X and Y and every property in between are undecidable, the formulation tacitly assumes that the inputs witnessing the undecidability never satisfy ``Y and not X.''  
A proof of Theorem~\ref{thm:undecidability} below can be found in Section~\ref{section:undecidability}.

\begin{restatable}{theorem}{undecidability}  
\label{thm:undecidability} For a given a Datalog sentence $\Phi$ over a signature containing at most binary relation symbols, it is undecidable whether 
 \begin{enumerate}
     \item simultaneously: {(a)} $\Phi$ is logically equivalent to a monadic Datalog sentence, {(b)} $\fm(\Phi)$ is the CSP of a finite structure,  {(c)} $\fm(\Phi)$ is the age of a finitely bounded-homogenizable structure, and {(d)} $\fm(\Phi)$ is the age of a reduct of a finitely bounded homogeneous Ramsey structure, or
     \item  none of the following is true:  {(e)} $\Phi$ is  logically equivalent to a guarded monotone SNP sentence, {(f)} $\fm(\Phi)$ is  the CSP of an $\omega$-categorical structure, or {(g)} $\fm(\Phi)$ is  the age of an $\omega$-categorical structure.
 \end{enumerate}  
\end{restatable}

Corollary~\ref{cor:extract} extracts the statement originally announced in the abstract.
\begin{corollary} \label{cor:extract} The question whether a given SNP sentence defines the age of a (finitely bounded-) homogenizable structure is undecidable.  
    The statement remains true even if the SNP sentence comes from the Datalog fragment and uses at most binary relation symbols.
\end{corollary}
 
We remark that we do not know whether the statement of Theorem~\ref{thm:undecidability} holds for any further restrictions to the set of instances, e.g., for universal first-order sentences, but we suspect that this is the case.

\medskip 
\noindent\textbf{Question 5:} Does the WHP remain undecidable even when its instances are restricted to universal first-order sentences?
\medskip

\subsection{Wider research context}

We now place our results within a wider research context.
An important special case of finitely bounded homogeneous structures are  \emph{stable} finitely homogeneous structures, which are precisely the unions of chains of finite homogeneous structures~\cite{lachlan1986homogeneous}.
Lachlan showed that the problem of testing if a given finite set of bounds over a finite relational signature specifies a stable amalgamation class is decidable~\cite[Theorem 6.2]{lachlan1997stable}; this result relies on the theory of stable finitely homogeneous structures developed by Cherlin, Harrington, and Lachlan~\cite{CHERLIN1985103,knight2006shrinking}.
The ADP generalizes Lachlan's decision problem from stable amalgamation classes to arbitrary finitely bounded amalgamation classes, where no structure theory is known. 

A different but related decision problem was presented at ICM’86~\cite{lachlan1986homogeneous}; we will refer to it as \emph{Lachlan's meta problem}.
Here, the task is to decide whether, given a pair $(\mathcal{A},\mathcal{B})$ of finite sets of finite structures over a finite relational signature $\tau$, there exists a countable homogeneous $\tau$-structure embedding every structure from $\mathcal{A}$ and none of the structures from $\mathcal{B}$.
Note that, if $\mathcal{B}$ is a YES-instance of the ADP, then $(\mathcal{A},\mathcal{B})$ is a YES-instance of Lachlan's meta problem if and only if $\mathcal{A} \subseteq \Forb_{e}(\mathcal{B})$.
However, a YES-instance $(\mathcal{A},\mathcal{B})$ of Lachlan's meta problem might be witnessed by uncountably many different homogeneous structures, which might not correspond to YES-instances of the ADP.
In Lachlan's words, his meta problem ``provides a way of giving a precise meaning to the problem of classification of arbitrary finitely homogeneous structures, or more exactly the problem of determining whether such a classification is possible''~\cite{cherlin1993combinatorial}.     
In contrast, the ADP is a direct expression of the classification problem for finitely bounded homogeneous structures as a computational decision problem.
Lachlan~\cite{lachlan1986homogeneous} conjectured that his meta problem is decidable; to the best of our knowledge, his conjecture has only been verified in very few cases, e.g., for a certain subset of instances in the special case of directed graphs~\cite{Latka_1994}. 

We  remark that Lachlan's meta problem has connections to the problem of testing whether a given class of finite structures specified by finitely many positive and negative constraints forms a \emph{Well Quasi-Order} (WQO) with respect to embeddability. 
This was observed by Cherlin~\cite{cherlin1993combinatorial,cherlin2011forbidden} during the classification of countable homogeneous directed graphs. 
Investigating the decidability of the WQO problem in special cases has later become a standalone subject~\cite{cherlin2000minimal,atminas2017wqo,mcdevitt2021atomicity,ironmonger2024decidability};  
we are not aware of any obvious connection between the WQO problem and the ADP.
The other problem whose decidability in concrete settings is often studied in these works is \emph{atomicity}, which corresponds to testing the AP restricted to empty intersections, commonly known as the \emph{Joint Embedding Property} (JEP).
We briefly discuss the difference between the JEP and the AP in Section~\ref{section:past}.

In theoretical computer science, the amalgamation property has on several occasions been considered/rediscovered as a useful structural restriction leading to guarantees of decidability or good runtime complexity.
In particular, there is an active branch of research focused on studying the complexity of CSPs parametrized by reducts of finitely bounded homogeneous structures.
It is believed that such problems are well-behaved from the algorithmic perspective, and that fundamental properties such as solvability in Datalog might be decidable.
At the same time, recognizing the structures whose CSPs are studied within this research programme turns out to be a very difficult problem, as demonstrated by our results on the ADP; we refer the reader to~\ref{subs:connections} for more information.
Our work on the WHP (Theorem~\ref{thm:undecidability}) shows that even a small change to the ADP can already lead to undecidability.
Admittedly, the theorem merely shows that SNP sentences are an exceptionally bad choice of an input to the WHP.
Nevertheless, this type of an input is often considered in areas at the intersection between model-theory and theoretical computer science, in particular in connection with CSPs; see, e.g.,~\cite{mottet2024promise,atserias2016non}. 

Besides CSPs, our results are also relevant to other related areas of theoretical computer science such as verification of database-driven systems~\cite{bojanczyk2013verification}, sets with atoms~\cite{clemente2015reachability}, or description logics with concrete domains~\cite{lutz2007tableau,baader2022using,borgwardt2024precise}. 
In all these areas, solving concrete instances of the ADP corresponds to finding concrete examples of structures to which the theory developed there is applicable.

\subsection{Connections to automata theory}
\begin{figure} 
     \centering
      \includegraphics[width=0.75\linewidth]{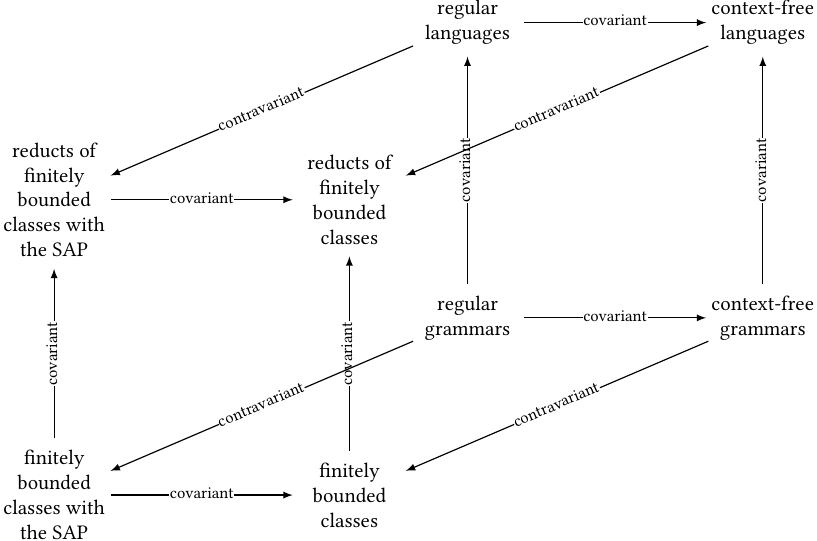} 
     \caption{Several functorial encodings relevant in the context of the proof of Theorem~\ref{thm:undecidability}. The edge-labels provide information about their variance with respect to containment.}
     \label{fig:inclusion_diagram}
 \end{figure}

In the proof of Theorem~\ref{thm:undecidability}, we encode context-free grammars and deterministic finite automata into finitely bounded classes.
In the case of deterministic finite automata, and also in the case of regular grammars, this can be done so that the resulting class has the SAP (even the FAP).
We invite the reader to view at Figure~\ref{fig:inclusion_diagram}, which contains a lattice of several encodings relevant in the context of the proof of Theorem~\ref{thm:undecidability}.
These encodings behave functorially with respect to containment, and the labels on the edges describe their variance.

The contravariance in some of the cases explains why Theorem~\ref{thm:inside_out} does not immediately yield the undecidability of the ADP even though the containment of regular languages in context-free languages is undecidable.
Our methods only provide an efficient reduction from the containment of context-free languages in regular languages, which is not harder than the containment between regular languages~\cite{asveld2000inclusion}.

Our results indicate that a more general version of this phenomenon might exist on the level of reducts of finitely bounded classes with the SAP.
The general message we want to convey is that the relationship between finitely bounded classes with the SAP and arbitrary finitely bounded classes shares similarities with the relationship between regular grammars and context-free grammars. 
A key difference is that the regularity of context-free grammars can be tested in linear time, while the problem of testing the SAP for finitely bounded classes is \TWONEXPTIME-hard (Corollary~\ref{cor:hardness}).

\subsection{Organisation of the article}
 
Section~\ref{section:preliminaries} provides some basic notions from logic necessary for the presentation of the proofs of our results.
Section~\ref{section:classification} provides some information about the ADP over binary signatures.
Section~\ref{section:inside_out} contains the proofs of   Theorem~\ref{thm:inside_out}, Theorem~\ref{thm:SAP_lemma}, and Theorem~\ref{thm:inside_out_2}.
Our proofs of these results are elementary and require very little external knowledge. 
 Section~\ref{section:cont_hard} contains the proof of Theorem~\ref{thm:containment_hard}. The proof is by a polynomial-time reduction from the 2-exponential square tiling problem.
The idea of the proof is inspired by the proof of \TWONEXPTIME-hardness of the containment problem for the logic MMSNP from~\cite{bourhis2016containment}. 
Section~\ref{section:ramsey} contains the proof of Theorem~\ref{theorem:ramsey}. The proof is by a direct application of our results and a combinatorial tool known as the canonisation lemma~\cite{bodirsky_pinsker_tsankov,bodirsky_pinsker_ramsey_canonical}.
The idea is inspired by the proof of decidability of the containment for classes of finite structures definable in the logic MMSNP from~\cite{bodirsky2021proof} and its generalization to the logic GMSNP in~\cite{barsukov2025containment}.
Section~\ref{section:undecidability} contains the proof of Theorem~\ref{thm:undecidability}. The proof is by a polynomial-time reduction from the problem of testing the regularity of context-free languages.  
\ref{subs:connections} is an extension of the introduction explaining the intricate connections to the field of infinite-domain constraint satisfaction; it is not essential for the rest of the article.

\section{Preliminaries} \label{section:preliminaries}
%
%
The set $\{1,\dots,n\}$ is denoted by $[n]$, and we use the bar notation for tuples.
We extend the usual containment relation on sets to tuples by ignoring the ordering on the entries. E.g., we might write $x \in \bar{t}$ if $x$ appears in an entry of a tuple $\bar{t}$.  
 
\subsection{Relational structures.}
A (\emph{relational}) \emph{signature} $\tau$ is a set of \emph{relation symbols}, each $R\in\tau$ with an associated natural number called \emph{arity}.
We say that $\tau$ is \emph{binary} if it consists of symbols of arity $\leq 2$.
A (\emph{relational}) \emph{$\tau$-structure} $\struct{A}$ consists of a set $A$ (the \emph{domain}) together with the relations $R^{\struct{A}}\subseteq A^{k}$ for each $R\in \tau$ with arity $k$.
An \emph{expansion} of $\struct{A}$ is a $\sigma$-structure $ \struct{B}$ with $A=B$ such that $ \tau\subseteq \sigma$, $R^{\struct{B}}=R^{\struct{A}}$ for each relation symbol $R\in \tau$. Conversely, we call $\struct{A}$ a \emph{reduct} of $\struct{B}$ and denote it by $\struct{B}|^{\tau}$.
These two notions naturally extend to classes of structures over a common signature.

Let $\struct{A}$ be a $\tau$-structure.
The \emph{substructure} of $\struct{A}$ on a subset $B\subseteq A$ is the $\tau$-structure $\struct{B}$ with domain $B$ and relations $R^{\struct{B}}=R^{\struct{A}}\cap B^k$ for every $R\in \tau$ of arity $k$.
The \emph{factor} of $\struct{A}$ through an equivalence relation $E\subseteq A^2$ is the $\tau$-structure $\struct{A}/{E}$ with domain $A/E$ and relations $R^{\struct{A}/{E}}= q_E(R^{\struct{A}})$, where $q_E$ denotes the factor map $x\mapsto [x]_E$.  
We say that $E$ is a \emph{relational congruence} on $\struct{A}$ if the definitions of the corresponding relations of $\struct{A}/E$ do not depend on the choices of the representatives of the equivalence classes of $E$:  for every $R\in \tau$ and all tuples $\bar{t}$ we have $q_E(\bar{t})\in R^{\struct{A}/{E}}$ if and only if $\bar{t}\in R^{\struct{A}}$. 
%
%

The \emph{union} of two $\tau$-structures $\struct{A}$ and $\struct{B}$ is the $\tau$-structure $\struct{A}\cup \struct{B}$ with domain $A\cup B$ and relations of the form $R^{\struct{A}\cup \struct{B}}\coloneqq R^{\struct{A}}\cup R^{\struct{B}}$ for every $R\in \tau$.
A \emph{disjoint union} of $\struct{A}$ and $\struct{B}$ is a union of two copies of $\struct{A}$ and $\struct{B}$ with disjoint domains.

 A \emph{homomorphism} $h\colon \struct{A} \rightarrow \struct{B}$ for $\tau$-structures $\struct{A}$ and $\struct{B}$ is a mapping $h\colon  A\rightarrow B$ that \emph{preserves} each relation of $\struct{A}$, i.e., if $ \bar{t} \in R^{\struct{A}}$ for some $k$-ary relation symbol $R\in \tau$, then $h(\bar{t})\in R^{\struct{B}}$.
We write $\struct{A} \rightarrow \struct{B}$ if $\struct{A}$ maps homomorphically to $\struct{B}$.
For a structure $\struct{A}$ over a finite relational signature $\tau$, we define $\CSP(\struct{A})$ as the class of all finite  $\tau$-structures which homomorphically map to $\struct{A}$.

An \emph{embedding} is an injective homomorphism $h\colon \struct{A} \rightarrow \struct{B}$ that additionally satisfies the following condition: for every $k$-ary relation symbol $R\in \tau$ and $\bar{t}\in A^{k}$ we have $h(\bar{t})\in R^{\struct{B}}$ only if $\bar{t}\in R^{\struct{A}}.$
For structures $\struct{A}$ and $\struct{B}$, we denote by $\binom{\struct{B}}{\struct{A}}$ the set of all embeddings of $\struct{A}$ into $\struct{B}$.
The \emph{age} of $\struct{A}$, denoted by $\Age(\struct{A})$, is the class of all finite structures which embed to $\struct{A}$. 
An \emph{isomorphism} is a surjective embedding. Two structures $\struct{A}$ and $\struct{B}$ are \emph{isomorphic} if there exists an isomorphism from $\struct{A} $ to $\struct{B}$.  

An \emph{automorphism} is an isomorphism from $\struct{A}$ to $\struct{A}$.  
The \emph{orbit} of a tuple $\bar{t}\in A^{k}$ in $\struct{A}$ is the set $\{g(\bar{t}) \mid g \text{ is an automorphism of }\struct{A}\}.$ 
A countable structure $\struct{A}$ is \emph{$\omega$-categorical} if, for every $k\geq 1$, there are finitely many orbits of $k$-tuples in $\struct{A}$.
Every homogeneous structure in a finite relational signature is $\omega$-categorical, and so are the reducts of such structures.
The following lemma can be shown by a  standard compactness argument, e.g., using K\H{o}nig's tree lemma.

\begin{lemma}[Lemma~4.1.7 in~\cite{Book}] \label{lemma:compactness} Let $\struct{A}$ and $\struct{B}$ be countable relational structures such that $\struct{B}$ is $\omega$-categorical.
Then $\struct{A}$ embeds into $\struct{B}$ if and only if every finite substructure of $\struct{A}$ embeds into $\struct{B}$.
\end{lemma}

\subsection{Fra\"{i}ss\'{e} and structural Ramsey theory} \label{section:fraisse}
Let $\mathcal{K}$ be a class of finite structures in a finite relational signature $\tau$ closed under isomorphisms and substructures.
An \emph{amalgamation diagram} for $\mathcal{K}$ is a pair of structures $\struct{B}_1,\struct{B}_2\in \mathcal{K}$ whose substructures on $B_1\cap B_2$ are identical.
An amalgamation diagram is \emph{one-point} if $|B_1\setminus B_2| = |B_2\setminus B_1|=1$.
Per convention, in a one-point amalgamation diagram $\struct{B}_1,\struct{B}_2$, we denote by $b_1$ and $b_2$ the unique elements contained in $B_1\setminus B_2$ and $B_2\setminus B_1$, respectively.
An \emph{amalgam} for an amalgamation diagram $\struct{B}_1,\struct{B}_2\in \mathcal{K}$ is a structure $\struct{C}\in \mathcal{K}$ for which there are embeddings $f_1\colon \struct{B}_1\rightarrow \struct{C}$ and $f_2\colon \struct{B}_2\rightarrow \struct{C}$
such that $f_1|_{B_1\cap B_2} = f_2|_{B_1\cap B_2}.$ 
Then $\mathcal{K}$ has the \emph{amalgamation property} (AP) if every amalgamation diagram for $\mathcal{K}$ has an amalgam in $\mathcal{K}$.  
The \emph{strong} version of the AP (SAP) is when the amalgam can always be chosen so that $f_1(B_1) \cap f_2(B_2) = f_1(B_1\cap B_2).$ 
Since $\mathcal{K}$ is closed under isomorphisms and substructures, without loss of generality, we may always assume that $f_1(B_1) \cup f_1(B_2) \subseteq B_1\cup B_2$ in the case of AP and $f_1(B_1) \cup f_1(B_2) = B_1\cup B_2$ in the case of SAP.
For a class $\mathcal{K}$ closed under isomorphisms and substructures, the AP is implied by the property of being closed under unions, also called \emph{free amalgams}.
In this case we say that $\mathcal{K}$ has the \emph{Free Amalgamation Property} (FAP).

Countable homogeneous structures arise as limit objects of well-behaved classes of finite structures in the sense of Fra\"{i}ss\'{e}'s theorem.
In the present article, we only need the following restriction of the theorem to finite relational signatures.
\begin{theorem}[Fra\"{i}ss\'{e}'s theorem, special case] \label{theorem:fraisse_2} For a class $\mathcal{K}$ of finite structures in a finite relational signature $\tau$, the following are equivalent:
\begin{itemize}
    \item $\mathcal{K}$ is closed under isomorphisms, substructures, and has the AP;
    \item $\mathcal{K}$ is the age of an up to isomorphism unique countable homogeneous $\tau$-structure; whenever such a structure exists, we refer to it as the \emph{Fra\"{i}ss\'{e} limit} of $\mathcal{K}$.
\end{itemize}
 
\end{theorem}

The literature usually refers to classes $\mathcal{K}$ satisfying the two equivalent conditions of Fra\"{i}ss\'{e}'s theorem as \emph{amalgamation classes}; such classes are \emph{strong} if the even have the SAP.
For finitely bounded classes, the closure under isomorphisms and substructures is trivially true, and hence finitely bounded amalgamation classes are simply finitely bounded classes with the AP.
By Theorem~\ref{theorem:fraisse_2}, every countable homogeneous structure is uniquely described by its age up to isomorphism.
Consequently, every countable finitely bounded homogeneous structure is uniquely described by a finite set of bounds or, equivalently, by a universal first-order sentence.  

A homogeneous structure is called \emph{Ramsey} if its age has the \emph{Ramsey property} (RP).
The precise definition of the RP is not essential to the present article; we only include it for the sake of completeness. 
A class $\mathcal{K}$ of structures over a common signature $\tau$ has the RP if,  for all $\struct{A},\struct{B}\in \mathcal{K}$ and every $k\in \mathbb{N}$, there exists $\struct{C}\in \mathcal{K}$ such that, for every map $f\colon \binom{\struct{C}}{\struct{A}} \rightarrow [k]$, there exists $e\in \binom{\struct{C}}{\struct{B}}$ such that $f$ is constant on the set $\bigl\{e\circ u\;|\; u\in \binom{\struct{B}}{\struct{A}}\bigr\} \subseteq \binom{\struct{C}}{\struct{A}}$.  
There is a large body of works focused on the study of homogeneous Ramsey structures; we refer the interested readers to~\cite{hubivcka2019all,hubivcka2025twenty}.
In the present article, we will only use this notion as a black-box.

Let $\struct{A}$ and $\struct{B}$ be two structures. 
A function $f\colon A \rightarrow B$ is called \emph{canonical} from $\struct{A}$  to $\struct{B}$ if, for every $m\in \mathbb{N}$, the componentwise action of $f$ induces a well-defined function from the orbits of
$m$-tuples in $\struct{A}$ to the orbits of $m$-tuples in $\struct{B}$.
Below we state a consequence of the so-called canonization lemma~(Theorem~5 in~\cite{bodirsky_pinsker_ramsey_canonical}).
Details about how to obtain Theorem~\ref{th:canonical_ramsey} from~\cite[ Theorem~5]{bodirsky_pinsker_ramsey_canonical} can be found in~\cite{barsukov2025containment}.
\begin{theorem}[\cite{bodirsky_pinsker_tsankov,bodirsky_pinsker_ramsey_canonical}]\label{th:canonical_ramsey}
Let $\struct{A}$ and $\struct{B}$ be structures over finite relational signatures such that $\struct{A}$ is homogeneous Ramsey and $\struct{B}$ is $\omega$-categorical. 
If there exists an embedding from a reduct $\struct{A}'$ of $\struct{A}$ to a reduct $\struct{B}'$ of $\struct{B}$, then there also exists an embedding from $\struct{A}'$ to $\struct{B}'$ that is canonical from $\struct{A}$ to $\struct{B}$. 
\end{theorem}

\subsection{First-order logic and SNP} \label{section:prelims_logic} 
We say that a first-order formula is \emph{$k$-ary} if it has $k$ free variables.
For a first-order formula $\phi$, we use the notation $\phi(\bar{x})$ to indicate that the free variables of $\phi$ are among $\bar{x}$.
This does not mean that the truth value of $\phi$ depends on each entry in $\bar{x}$.
We assume that equality $=$ as well as the nullary predicate symbol $\bot$ for falsity are 
always available when building first-order formulas. 
Thus, \emph{atomic $\tau$-formulas}, or \emph{$\tau$-atoms} for short, over a relational signature $\tau$ are of the form $\bot$, $(x=y)$, and $R(\bar{x})$ for some $R\in \tau$ and a tuple $\bar{x}$ of first-order variables matching the arity of $R$.

A \emph{universal} first-order $\tau$-sentence is of the form $\forall \bar{x} \ldotp \phi(\bar{x})$ for a quantifier-free $\tau$-formula $\phi$.
Let $\Phi$ be a universal first-order $\tau$-sentence whose quantifier-free part $\phi$ is in CNF, i.e., a conjunction of \emph{clauses}, which are disjunctions of possibly negated $\tau$-atoms. 
We call $\Phi$ \emph{Horn} if every clause of $\phi$ is Horn, i.e., contains at most one positive disjunct.   
For a clause $\phi_i$ of $\phi$, we define the \emph{Gaifman graph} of $\phi_i$ as the undirected graph whose vertex set consists of the variables appearing in some atom of $\phi_i$ and where two distinct variables $x,y$ form an edge if and only if they appear jointly in a negative atom of $\phi_i$. 
A clause $\phi_i$ of $\phi$ is \emph{complete} (\emph{connected}) if the Gaifman graph of $\phi_i$ is complete (connected).
The crucial part of this definition is that all variables in a clause count towards the vertices of the Gaifman graph but only the negative atoms count towards its edges; e.g., $\neg E(x,y)\vee \neg E(x,z) \vee \neg E(y,z)$ is complete while $\neg E(x,y)\vee \neg E(x,z) \vee E(y,z)$ is connected but not complete.
We call $\Phi$ \emph{complete} (\emph{connected}) if every clause of $\phi$ is \emph{complete} (\emph{connected}). 
It is a folklore fact that, if $\Phi$ is complete (connected), then $\fm(\Phi)$ is preserved by (disjoint) unions. 
 
 An \emph{SNP $\tau$-sentence} is a second-order sentence $\Phi$ of the form 
$
\exists X_1,\dots, X_n \forall \bar{x} \ldotp \phi
$
where $\phi$ is a quantifier-free formula in CNF over $\tau\cup \{X_1,\dots, X_n \}$.
We call $\Phi$ \emph{monadic} if $X_i$ is unary for every $i\in [n]$; \emph{monotone} if $\phi$ does not contain any positive $\tau$-atoms (in particular no positive equality atoms); and \emph{guarded} if, for every positive atom $\beta$ in a clause of $\phi$ there exists a negative atom $\alpha$ in the same clause of $\phi$ containing all variables of $\beta$.  
The Monadic Monotone and the Guarded Monotone fragments of SNP are denoted by MMSNP and GMSNP, respectively. 
Up to a minor syntactic modification, GMSNP is a proper generalization of MMSNP~\cite{bienvenu2014} (see also~\cite[Section~1.1]{barsukov2025containment}).
A typical example of a class of structures definable by a GMSNP sentence is the class of all graphs that can be edge-2-colored while avoiding monochromatic triangles:
\begin{align*} 
  \exists B, R\, \forall x,y,z &\big(  \neg E(x,y) \vee   \neg  E(y,z) \vee \neg E(z,x) \vee \neg B(x,y) \vee   \neg  B(y,z) \vee \neg B(z,x)   
    \big)  \\
     {} \wedge \,  &\big(   \neg  E(x,y) \vee  \neg E(y,z) \vee \neg E(z,x) \vee  \neg R(x,y) \vee  \neg R(y,z) \vee \neg  R(z,x)   
    \big)   \\ 
   {} \wedge \,  &\big( \neg E(x,y) \vee B(x,y) \vee R(x,y) \big) \wedge \big( \neg E(x,y) \vee \neg B(x,y) \vee \neg R(x,y) \big).
\end{align*}   
In MMSNP, we can instead define the class of all graphs that can be vertex-2-colored while avoiding monochromatic triangles.   
Note that the monotonicity axiom prevents us from forcing $E$-edges to be undirected but this does not matter for our demonstrative purposes.  
The following theorem was proved in~\cite{bodirsky2020asnp}, except that the authors did not fully utilize the auxiliary result from~\cite{hubivcka2019all} guaranteeing the existence of a Ramsey expansion by a generic linear order; see~\cite{barsukov2025containment} for more details.
\begin{theorem}[\cite{bodirsky2020asnp}] \label{theorem:gmsnp_homogenizable} 
Every GMSNP sentence is equivalent to a finite disjunction of connected GMSNP sentences. 
Every connected GMSNP sentence defines the age and the CSP of a reduct of a finitely bounded homogeneous Ramsey structure.
\end{theorem}

The notions of Horn, connected, and complete sentences easily transfer to SNP sentences viewed as universal sentences in an extended signature.
The monotone Horn fragment of SNP is commonly known as the logic programming language \emph{Datalog}.
When we say that a Datalog sentence $\Phi$ \emph{solves} the CSP of a structure $\struct{B}$, we simply mean that $\fm(\Phi)=\CSP(\struct{B})$.
This is consistent with the usual definition of the complementary class being definable in Datalog viewed as an extension of the existential positive fragment of first-order logic by formation rules whose semantics is defined via inflationary fixed-points of definable operators (cf.~\cite{feder1998computational}).

\subsection{Formal languages}

As usual, the \emph{Kleene plus} and the \emph{Kleene star} of a finite set of symbols $\Sigma$, denoted by $\Sigma^{+}$ and $\Sigma^{*}$, are the sets of all finite words over $\Sigma$ of lengths $\geq 1$ and $\geq 0$, respectively.

A \emph{context-free grammar} (CFG) is a $4$-tuple $\fancyfont{G}=(N,\Sigma,P,S)$ where 
$N$ is a finite set of \emph{non-terminal symbols}, $\Sigma$ is a finite set of \emph{terminal symbols}, $P$ is a finite set of \emph{production rules} of the form $A \rightarrow w$ where $A\in N$ and $w \in (N \cup \Sigma)^{+}$, $S \in N$ is the \emph{start symbol}.
For $u,v\in (N\cup \Sigma)^{+}$ we write $u \rightarrow_{\fancyfont{G}} v $ if there are $x,y\in(N \cup \Sigma)^{+} $ and $(A \rightarrow w)\in P$ such that $u=xAy$ and $v=xwy$.
The \emph{language of $\fancyfont{G}$} is $L(\fancyfont{G}) \coloneqq \{ w\in \Sigma^{+} \mid S \rightarrow_{\fancyfont{G}}^{\ast} w \},$ where $\rightarrow_{\fancyfont{G}}^{\ast}$ denotes the transitive closure of $\rightarrow_{\fancyfont{G}}$.
Note that with this definition the \emph{empty word}, i.e., the word $\varepsilon$ of length $0$, can never be an element of $L(\fancyfont{G})$; some authors use a modified definition that also allows rules that derive $\varepsilon$, but for our purposes the difference is not essential. 

A CFG is called \emph{(left-)regular} if its production rules are always of the form $A\rightarrow a$ or $A\rightarrow Ba$ for non-terminal symbols $A,B$ and a terminal symbol $a$.
For a finite set $\Sigma$, we call a set $L\subseteq \Sigma^{+}$ \emph{regular} if it is the language of a regular grammar with terminal symbols $\Sigma$. 
As an example, consider the CFG $\fancyfont{G}$ with a single terminal symbol $a$, non-terminal symbols $S,A,B,C$, and production rules $S\rightarrow a$, $S\rightarrow aa$, $S\rightarrow aaa$, $S\rightarrow Aa$, $A\rightarrow Ba$, $B\rightarrow Ca$, $C\rightarrow Ca$, and $C\rightarrow a$.
Clearly, the grammar $\fancyfont{G}$ is not regular.
However, its language $L(\fancyfont{G})=\{a\}^{+}$ is regular because it is also the language of the regular grammar $\fancyfont{G}'$ with the production rules $S\rightarrow Sa$ and $S\rightarrow a$.

 A \emph{deterministic finite automaton} (DFA) is a $5$-tuple $\fancyfont{A}=(Q,\Sigma,\delta,q_0,F)$ where $Q$ is a finite set of \emph{states}, $\Sigma$ is a finite set of \emph{input symbols}, $\delta:Q\times \Sigma \rightarrow Q$ is a \emph{transition function}, $q_0\in Q$ is a distinguished \emph{starting state}, and $F\subseteq Q$ is a distinguished set of \emph{final states}. 
The \emph{language} of $\fancyfont{A}$ is $L(\fancyfont{A})\coloneqq \{a_1\dots a_n \in \Sigma^{+} \mid \delta(a_n,\dots\delta(a_{2},\delta(a_{1},q_0))\dots)\in F\}$.
Note that, as in the case of CFGs, the empty word $\varepsilon$ can never be an element of $L(\fancyfont{A})$ according to the present definition.
Let $\Sigma$ be a finite set of symbols and $L$ a subset of $\Sigma^{+}$. 
The \emph{Myhill-Nerode equivalence relation} on $\Sigma^{*}$, denoted by $\sim_L$, is defined by $w_1 \sim_L w_2 $ if there is no $w\in \Sigma^{*}$ such that $|\{w_1 w,w_2 w\}\cap L|=1$.  
The following correspondence is well-known.
\begin{theorem}[Myhill-Nerode] \label{theorem:myhil_nerode} For every finite set $\Sigma$ and every $L\subseteq \Sigma^+$, the following are equivalent:
\begin{itemize}
    \item $L$ is regular; 
    \item $L$ is accepted by a DFA;
    \item $\sim_L$ has finitely many classes. 
\end{itemize}
\end{theorem}
 

\section{The ADP Restricted to Binary Signatures} \label{section:classification}

It is known that the question whether a finitely bounded class has the AP can be tested algorithmically in the case where the signature is binary~\cite{lachlan1986homogeneous}.
This decidability result is based on the following observation. 

\begin{proposition}[\cite{lachlan1986homogeneous}]   \label{prop:one_point_AP} A class of finite relational $\tau$-structures that is closed under isomorphisms and substructures has the AP if and only if it has the AP restricted to one-point amalgamation diagrams. 
\end{proposition} 

As a consequence of Proposition~\ref{prop:one_point_AP}, if a finitely bounded class over a binary signature does not have the AP, then the size of a smallest counterexample to the AP is polynomial in the size of the set of bounds~\cite{bodirsky2020asnp}.
Such a counterexample can be non-deterministically guessed and verified using a \textup{\textsf{coNP}}-oracle, which places the problem at the second level of the polynomial hierarchy (Theorem~15 in~\cite{baader2022using}).
Note that this upper bound only applies to the setting where the input is specified by a set of bounds.
If the input is specified by a universal first-order sentence, then it might be the case that a smallest set of bounds witnessing finite boundedness is exponentially larger.
Consequently, the algorithm from \cite{bodirsky2020asnp} only gives us a relatively weak upper bound for the case where the inputs are specified by universal sentences.
	\begin{proposition} \label{cor:conexptime}  Let $\Phi$ be a universal sentence over a finite binary relational signature $\tau$.
			If $\fm(\Phi)$ does not have the AP, then the size of a smallest counterexample to the AP is at most single-exponential in the size of $\Phi$.
			Consequently, the question whether $\fm(\Phi)$ has the AP is decidable in \coNEXPTIME. 
\end{proposition}
\begin{proof} 
	Suppose that $\fm(\Phi)$ does not have the AP. 
	By Proposition~\ref{prop:one_point_AP}, we may assume that a counterexample to the AP for $\fm(\Phi)$ is a one-point amalgamation diagram formed by two structures $\struct{B}_1,\struct{B}_2 \in \fm(\Phi)$.  
    Let $k$ be the number of variables in $\Phi$.
    We define $\mathcal{B}$ as the set of all $\tau$-structures $\struct{A}$ such that $\struct{A}\centernot{\models}\Phi$, $|A|\leq k$, and $A=[|A|]$.
    Clearly, $\Forb_e(\mathcal{B})=\fm(\Phi)$. 
    Since the signature of $\Phi$ is binary, the size of $\mathcal{B}$ is at most $2^{|\tau|  k^2}$, i.e., single-exponential in the size of $\Phi$.
    By the proof of Theorem~4 in \cite{bodirsky2020asnp}, we may assume that the size of $\struct{B}_1$ and $\struct{B}_2$ is polynomial in the size of $\mathcal{B}$.
    Hence, we may assume that it is at most single-exponential in the size of $\Phi$. 
	
	Now, we must verify that $\struct{B}_1,\struct{B}_2 \in \fm(\Phi)$.
	This can be done in time exponential in the size of $\Phi$ simply by evaluating the quantifier-free part of $\Phi$ on all possible inputs.
	Subsequently, we must verify that no amalgam $\struct{C}\in \fm(\Phi) $ of $\struct{B}_1$ and $\struct{B}_2$ can be obtained   either by identifying $b_1$ and $b_2$, or by adding   $(b_1,b_2)$ or $(b_2,b_1)$ to some relations of $\struct{B}_1\cup \struct{B}_2$.
	This can also be done in time exponential in the size of $\Phi$ because only single-exponentially many structures $\struct{C}$ need to be checked. In sum, the existence of a counterexample can be tested non-deterministically in exponential time, which is what we had to show.
\end{proof}

The upper bound provided by Proposition~\ref{cor:conexptime} is not entirely unreasonable since a smallest counterexample to the AP might be of size exponential in the size of the input sentence even for binary signatures.
Such situations arise, e.g., in the proof of the next proposition.
We present Proposition~\ref{prop:binary_intro} together with a full proof as a warm-up for the more technically involved arguments in the upcoming sections. 

\begin{proposition} \label{prop:binary_intro} 
Given a universal sentence $\Phi$ over a finite binary relational signature, the question whether $\fm(\Phi)$ has the AP is $\Pi^p_3$-hard.
\end{proposition}
\begin{proof} We reduce from $\forall \exists \forall $-DNF, a basic complete problem for the complexity class $\Pi^p_3$~\cite{stockmeyer1976polynomial}.
Consider a general instance of $\forall \exists \forall $-DNF, which is of the form
\begin{align}
    \forall x_1,\dots, x_k\exists x_{k+1},\dots, x_{\ell} \forall x_{\ell+1},\dots, x_{m}  \ldotp \phi,  \label{eq:instance}
\end{align}
where $\phi$ is a disjunction of conjunctions of possibly negated propositional variables.

We define the signature $\tau$ as follows.
For every $i\in [m]$,  $\tau$ contains the symbol $X_i$, which is binary for $i\in [\ell]\setminus [k]$ and unary otherwise.
In addition, $\tau$ also contains the binary symbol $P$ and the two unary symbols $L$ and $R$.
We set
\[\Phi \coloneqq \forall x,y_1,y_2  \big( L(y_1) \wedge R(y_2) \wedge P(x,y_1) \wedge P(x,y_2) \Rightarrow   \phi'\big),\]
where $\phi'$ is the $\tau$-formula obtained from $\phi$ by the following syntactical replacement of propositional variables by $\tau$-atoms.
For every $i\in [m]$ we replace each instance of the variable $x_i$ by:
\begin{itemize}
    \item $X_i(y_1)$ if $i\in [k]$,
    \item $X_i(y_1,y_2)$ if $i\in [\ell]\setminus [k]$,
    \item $X_i(x)$ if $i\in [m]\setminus [\ell]$.
\end{itemize}

What we just did is we transformed $\phi$ into a universal first-order $\tau$-sentence $\Phi$ with the property that testing the satisfiability of $\phi$ translates to testing whether a particular one-point amalgamation diagram $\struct{B}_1,\struct{B}_2\in \fm(\Phi)$  has an amalgam in $\fm(\Phi)$.
In addition, the said one-point amalgamation diagram is the only potential obstruction to the AP for $\fm(\Phi)$.
We have that either $\phi$ is not satisfiable and consequently $\fm(\Phi)$ does not have the AP, or $\struct{B}_1,\struct{B}_2$ and every other one-point amalgamation diagram for $\fm(\Phi)$ can be completed to an amalgam in $\fm(\Phi)$ by fixing 
mistakes stemming from specific evaluations of the quantifier-free part of $\Phi$ in $\struct{B}_1\cup \struct{B}_2$.
Morally, the unary predicates $L$ and $R$ mark the distinguished left and the right elements in one-point amalgamation diagrams, i.e., the unique elements of $B_1\setminus B_2$ and $B_2\setminus B_1$. Note that $\Phi$ is true in all structures where $L$ or $R$ interprets as an empty relation, which ensures that $\struct{B}_1,\struct{B}_2\in \fm(\Phi)$.
The binary predicate $P$ marks those pairs of variables $(y_1,y_2)$ where the substitution of $(b_1,b_2)$, $(b_2,b_1)$ automatically makes the quantifier-free part of $\Phi$ evaluate as true in $\struct{B}_1\cup \struct{B}_2$; this gives us control over the amalgamation mistakes.
The details are given below.  

``$\Rightarrow$'' Suppose that \eqref{eq:instance} is satisfiable. 
Let $\struct{B}_1,\struct{B}_2$ be an arbitrary one-point amalgamation diagram for $\fm(\Phi)$.
If $\struct{B}_1\cup \struct{B}_2\models \Phi$, then we are done because $\struct{B}_1\cup \struct{B}_2$ is an amalgam for $\struct{B}_1$ and $\struct{B}_2$.
So suppose that instead $\struct{B}_1\cup \struct{B}_2 \centernot{\models}\Phi$.
Consider an evaluation of the quantifier-free part of $\Phi$ witnessing the fact that $\struct{B}_1\cup \struct{B}_2 \centernot{\models}\Phi$.
By definition, $b_1$ and $b_2$ do not appear together in any relation of $\struct{B}_1\cup \struct{B}_2$ and $\struct{B}_1,\struct{B}_2\models\Phi$.
Thus, by the shape of $\Phi$, it must be the case that $x$ is assigned some element $b\in B_1\cap B_2$, $y_1$ is assigned $b_1$, and $y_2$ is assigned $b_2$ (or vice versa).
For $i\in \{1,2\}$, define $T_i\coloneqq \{j \in [k] \mid b_i\in X^{\struct{B}_i}_j\}.$ 
Now, for a fixed $i\in \{1,2\}$, consider the propositional assignment where, for every $j\in [k]$, the truth value of $x_j$ is set to \emph{true} if and only if $j\in T_i$.
By the assumption that \eqref{eq:instance} is satisfiable, there exists $T'_i \subseteq [\ell]\setminus [k]$ such that the following holds: if, for every $j\in [\ell]\setminus [k]$, we set 
$x_j$ to \emph{true} if and only if $j\in T'_i$, then every assignment of truth values to the remaining variables $x_{\ell+1},\dots,x_{m}$ satisfies the quantifier-free part of~\eqref{eq:instance}.
We obtain an amalgam $\struct{C} \in \fm(\Phi)$ for $\struct{B}_1$ and $\struct{B}_2$ by adding, for every  $i\in \{1,2\}$ and $j\in [\ell]\setminus [k]$, the pair $(b_i,b_{3-i})$ to $X^{\struct{B}_1\cup \struct{B}_2}_{j}$ if and only if $j\in T'_{i}$.

``$\Leftarrow$''  Suppose that \eqref{eq:instance} is not satisfiable. 
Then there exists $T\subseteq [k]$ together with an assignment of truth values to the variables $x_1,\dots, x_k$ witnessing the unsatisfiability of \eqref{eq:instance} such that precisely $\{x_i\mid i\in T\}$ are set to \emph{true} in the assignment.
We use $T$ to construct a one-point amalgamation diagram $\struct{B}_1,\struct{B}_2\in \fm(\Phi)$ with no amalgam in $\fm(\Phi)$.
For $i\in \{1,2\}$, we set 
$
B_i\coloneqq \{b_i\}\cup \{b_S \mid S\subseteq [m]\setminus [\ell]\}.
$
The relations are as follows. 
We have $L^{\struct{B}_1}=\{b_1\}$,  $R^{\struct{B}_1}=\emptyset$, $L^{\struct{B}_2}=\emptyset $,  $R^{\struct{B}_2}=\{b_2\}$. 
Moreover, for every $j\in T$, $S\subseteq [m]\setminus [\ell]$, and  $i\in S$, %
we have $b_1 \in X_j^{\struct{B}_1}$, $(b_S,b_1)\in P^{\struct{B}_1}$, $b_S \in X_i^{\struct{B}_1}$, $b_2 \in X_j^{\struct{B}_2}$, $(b_S,b_2)\in P^{\struct{B}_2}$, and $b_S \in X_i^{\struct{B}_2}$.  
There are no other tuples in the relations of $\struct{B}_1$ or $\struct{B}_2$. 
Note that $\struct{B}_1,\struct{B}_2\in \fm(\Phi)$ because $R^{\struct{B}_1}=\emptyset$ and $L^{\struct{B}_2}=\emptyset$.
Clearly, no amalgam for $\struct{B}_1$ and $\struct{B}_2$ can be obtained by identifying $b_1$ and $b_2$ because $L^{\struct{B}_1}=\{b_1\}$ and  $R^{\struct{B}_2}=\{b_2\}$.
By the assumption that \eqref{eq:instance} is not satisfiable, there is also no strong amalgam for $\struct{B}_1$ and $\struct{B}_2$ satisfying $\Phi$ because the pair $(b_1,b_2)$ cannot be present in any subset of the relations $X_{k+1}^{\struct{B}_1\cup \struct{B}_2}, \dots, X_{\ell}^{\struct{B}_1\cup \struct{B}_2}$. 
\end{proof} 

\subsection{Moving past binary signatures} \label{section:past}
  
Very little progress has been made on signatures that contain symbols of arities greater than $2$.
In particular, it is not known whether the ADP is decidable.
The scenario where this is not the case is not entirely unrealistic since the closely related JEP is already undecidable for finitely bounded classes of graphs~\cite{BraunfeldUndec}.
There is, however, one fundamental difference between the JEP and the AP, which makes it quite unlikely that there could be a simple proof of undecidability of the ADP.

Intuitively, it is rather easy to encode configurations of Turing machines into finitely bounded classes preserved by disjoint unions,  which trivially have the JEP; we only need to define them by connected sentences (cf.~\cite[Appendix~A]{gaifman1993undecidable}).
Having done that, to obtain a reduction from some undecidable problem to the problem of testing the JEP, we must only create artificial failures of the JEP from some non-trivial property of Turing machines.
In practice, it is more convenient to reduce from some ``predigested''  problem, e.g., the unrestricted tiling problem~\cite{BraunfeldUndec} or the universality problem for context-free grammars~\cite{schrottenloher2022universal}. 

In contrast, to encode configurations of Turing machines into finitely bounded free amalgamation classes using complete sentences, one needs to include extra padding predicates, whose involvement makes any further use of such an encoding hopeless.
One way to circumvent this problem is to only consider Turing machines with a suitably bounded runtime, in which case a substantial part of the encoding can be outsourced to the signature and the padding becomes less obtrusive. 
This is what we implicitly do in our proof of the \TWONEXPTIME-hardness of the ADP (Sections~\ref{section:inside_out} and~\ref{section:cont_hard}).
Theorem~\ref{theorem:ramsey} shows that, under some reasonable assumptions, our lower bound might in fact not be that far from an upper bound.

\section{The Inside-Out Correspondence} \label{section:inside_out} 

In this section, we prove three of the results announced in the first part of the introduction.
We start with the proof of Theorem~\ref{thm:inside_out}, which is restated below. 

\firstinsideout*
 \begin{proof} 
The idea is to merge $\Phi_1$ and $\Phi_2$ into a single sentence $\Phi$ such that the question whether $\struct{A}\in \fm(\Phi_2)|^{\tau}$ holds for a given $\struct{A}\in \fm(\Phi_1)|^{\tau}$ can be translated to the question whether a particular one-point amalgamation diagram for $\fm(\Phi)$ based on $\struct{A}$ has an amalgam in $\fm(\Phi)$.
The sentence $\Phi$ will also have the property that the one-point amalgamation diagrams for $\fm(\Phi)$ without any amalgam in $\fm(\Phi)$ correspond precisely to the structures in $ \fm(\Phi_1)|^{\tau}\setminus \fm(\Phi_2)|^{\tau}$.
To this end, we proceed using a padding technique similar to the one in the proof of Proposition~\ref{prop:binary_intro}. The prerequisite for this technique to work is that $\fm(\Phi_1)$ has the SAP; otherwise, there could be unintended failures of the SAP for $\fm(\Phi)$ and the proof would break down.
In the following, we give the details of the construction and verify the equivalence of the three items in the theorem.

Let $\sigma_1$ and $\sigma_2$ be the signatures that gather the remaining symbols in $\Phi_1$ and $\Phi_2$ (for convenience different from the symbols in $\tau$).
We denote the quantifier-free parts of $\Phi_1$ and $\Phi_2$ by $\phi_1(\bar{x}_1)$ and $\phi_2(\bar{x}_2)$, respectively.  
We first introduce two fresh unary symbols $L,R$ and a fresh binary symbol $P$. 
Then, we introduce the signature $\sigma'_2$ that contains, for each $R\in \sigma_2$ of arity $k$, a symbol $R'$ of arity $k+2$.
Next, we set  
\begin{align*}
 \Phi'_1 \coloneqq & \ \forall \bar{x}_1  \big(   \phi_1(\bar{x}_1)\vee  \bigvee\nolimits_{x\in \bar{x}_1} L(x) \vee R(x) \big), \\   
\Phi'_2\coloneqq & \ \forall y_1,y_2,\bar{x}_2    \big( L(y_1) \wedge R(y_2)    \wedge \bigwedge\nolimits_{x\in \bar{x}_2} P(y_1,x) \wedge P(y_2,x)     
   \\  &    {}  \Rightarrow   \bigvee\nolimits_{
     (x_1,x_2)\in (\bar{x}_2,y_1,y_2),\, \{x_1,x_2\}\neq \{y_1,y_2\}} \big(P(x_1,x_2)\wedge P(x_2,x_1) \big)  \\  &    {} \hspace{10em}
   \vee   \phi'_2(\bar{x}_2,y_1,y_2)       \vee\bigvee\nolimits_{x\in \bar{x}_2} L(x) \vee R(x)  
   \big),  
\end{align*} 
 where $\phi'_2$ denotes the formula obtained from $\phi_2$ by replacing each $\sigma_2$-atom $X(\bar{t})$ by the $\sigma'_2$-atom $X'(\bar{t},y_1,y_2)$.
Finally, we set $\Phi\coloneqq  \Phi'_1\wedge \Phi'_2 $ (and bring $\Phi$ into prenex normal form).

``$\eqref{item:inside_out2}\Rightarrow\eqref{item:inside_out3}$'' This direction is trivial.

``$\eqref{item:inside_out3}\Rightarrow\eqref{item:inside_out1}$'' 
For the first part, let $\struct{B}_1,\struct{B}_2$ be an arbitrary one-point amalgamation diagram for $\fm(\Phi_1)$.
Consider the $(\tau\cup \sigma_1\cup \sigma'_2\cup \{L,R,P\})$-expansions  $\struct{B}'_1,\struct{B}'_2$ of $\struct{B}_1,\struct{B}_2$ by empty relations, respectively, except that $(b_1,b_1)\in P^{\struct{B}'_1}$ holds for the unique element $b_1\in B_1\setminus B_2$.
Then $\struct{B}'_1,\struct{B}'_2\models \Phi'_1$ is inherited directly from $\struct{B}_1,\struct{B}_2\models \Phi_1$.
Moreover, $\struct{B}'_1,\struct{B}'_2\models \Phi'_2$ holds because $L$ interprets as the empty relation in both $\struct{B}'_1$ and $\struct{B}'_2$ (one could clearly also argue with $R$).
Hence, $\struct{B}'_1,\struct{B}'_2$ is a one-point amalgamation diagram for $\fm(\Phi)$.
Since $\fm(\Phi)$ has the AP, there exists an amalgam $\struct{C}'$ for $\struct{B}'_1$  and $\struct{B}'_2$ satisfying $\Phi$.
We have that $\struct{C}'$ is strong because $P^{\struct{B}'_1}=\{(b_1,b_1)\}$ while $P^{\struct{B}'_2} = \emptyset$.
Without loss of generality, $C' = B_1\cup B_2$. 
We define $\struct{C}$ as the $(\tau\cup \sigma_1)$-reduct of $\struct{C}'$.
We have $\struct{C}\models \Phi_1$ because both $L$ and $R$ interpret as the empty relation in $\struct{C}'$ and $(\sigma'_2\cup \{P\})$-atoms do not interact with $\Phi_1'$.
We conclude that $\fm(\Phi_1)$ has the SAP. 

For the second part, let $\struct{A}\in \fm(\Phi_1)|^{\tau}$ be arbitrary, and let $\struct{A}'$ be an arbitrary $(\tau\cup \sigma_1)$-expansion of $\struct{A}$ satisfying $\Phi_1$.
We define the one-point amalgamation diagram $\struct{B}_1,\struct{B}_2$ for $\fm(\Phi)$ as follows.
The domains of $\struct{B}_1$ and $\struct{B}_2$ are $A\cup \{b_1\}$ and $A\cup \{b_2\}$, respectively, and the relations are inherited from $\struct{A}'$, except that additionally $b_1\in L^{\struct{B}_1}$, $b_2\in R^{\struct{B}_2}$, and, for every $a\in A$, we have $(b_1,a)\in P^{\struct{B}_1}$ and $(b_2,a)\in P^{\struct{B}_2}$.
Since $L^{\struct{B}_1}=\{b_1\}$, $R^{\struct{B}_1} = \emptyset$, and $\struct{A}'\models\Phi_1$, we get that $\struct{B}_1\models \Phi'_1$.
Since $L^{\struct{B}_1}=\{b_1\}$ and $R^{\struct{B}_1} = \emptyset$, we get that $\struct{B}_1\models \Phi'_2$.
In sum, $\struct{B}_1\models \fm(\Phi)$.
Similarly we get that $\struct{B}_2\models \fm(\Phi)$. 
Hence, $\struct{B}_1,\struct{B}_2$ is a one-point amalgamation diagram for $\fm(\Phi)$.
Since $\fm(\Phi)$ has the AP, there exists an amalgam $\struct{C}$ for $\struct{B}_1$ and $\struct{B}_2$ satisfying $\Phi$.
This amalgam must be strong because $L^{\struct{B}_1}=\{b_1\}$ while $L^{\struct{B}_2} = \emptyset$.
Let $\struct{A}''$ be the $(\tau\cup \sigma_2)$-expansion of $\struct{A}$ such that, for every $X\in \sigma_2$, we have $\bar{t}\in X^{\struct{A}''}$ if and only if $(\bar{t},b_1,b_2)\in {X'}^{\struct{C}}$.
Since $\struct{C}\models \fm(\Phi)$, we have that $\struct{A}''\models \Phi_2$, and hence $\struct{A}\in \fm(\Phi_2)|^{\tau}$. 
We conclude that $\fm(\Phi_1)|^{\tau}\subseteq\fm(\Phi_2)|^{\tau}$. 

``$\eqref{item:inside_out1}\Rightarrow\eqref{item:inside_out2}$'' Let $\struct{B}_1,\struct{B}_2$ be a one-point amalgamation diagram for $\fm(\Phi)$.
 
First, suppose that $b_1,b_2 \notin L^{\struct{B}_1}\cup R^{\struct{B}_1}\cup L^{\struct{B}_2}\cup R^{\struct{B}_2}$.
We obtain $\struct{B}'_1$ and $\struct{B}'_2$ from $\struct{B}_1$ and $\struct{B}_2$ by taking their substructures on $B_1\setminus (L^{\struct{B}_1}\cup R^{\struct{B}_1})$ and $B_2\setminus (L^{\struct{B}_2}\cup R^{\struct{B}_2})$, respectively.
Let $\struct{B}''_1$ and $\struct{B}''_2$ be the $(\tau\cup \sigma_1)$-reducts of $\struct{B}'_1$ and $\struct{B}'_2$, respectively. 
By the definition of $\Phi'_1$, we get that $\struct{B}''_1,\struct{B}''_2\models \Phi_1$ because  $\struct{B}_1,\struct{B}_2\models \Phi'_1$.
Since $\fm(\Phi_1)$ has the SAP, there exists a strong amalgam $\struct{C}'$ for $\struct{B}''_1$ and $\struct{B}''_2$ satisfying $\Phi_1$.
Without loss of generality, we have $C'= B'_1\cup B'_2$. 
We obtain $\struct{C}$  by adding tuples to the relations of $\struct{B}_1\cup \struct{B}_2$ so that $(b_1,b_2),(b_2,b_1)\in P^{\struct{C}}$ and, for every $X\in \tau\cup \sigma_1$, we have  $\bar{t}\in X^{\struct{C}}$ if and only if $\bar{t}\in X^{\struct{C}'}$.
Note that, by the second line of the definition of $\Phi'_2$, the quantifier-free part of $\Phi'_2$ is trivially true on $\struct{C}$ for all assignments to the free variables whose range contains both $b_1$ and $b_2$.

Second, suppose that $\{b_1,b_2\} \cap (L^{\struct{B}_1}\cup R^{\struct{B}_1}\cup L^{\struct{B}_2}\cup R^{\struct{B}_2})\neq \emptyset$.
We obtain $\struct{A}$ by first taking the substructure of $\struct{B}_1$ (or equivalently $\struct{B}_2$) on $(B_1\cap B_2)\setminus (L^{\struct{B}_1}\cup R^{\struct{B}_1})$.
Since $\struct{B}_1\models \Phi$, we have $\struct{A}\in \fm(\Phi_1)$.
Let $\struct{A}'$ be the $\tau$-reduct of $\struct{A}$.
Since $\fm(\Phi_1)|^{\tau}\subseteq \fm(\Phi_2)|^{\tau}$, there exists a $(\tau\cup \sigma_2)$-expansion $\struct{A}''$ of $\struct{A}'$ satisfying $\Phi_2$.
We obtain $\struct{C}$ by adding tuples to the relations of $\struct{B}_1\cup \struct{B}_2$ so that $(b_1,b_2),(b_2,b_1)\in P^{\struct{C}}$ and, for every $X\in \tau\cup \sigma_2$, we have $(\bar{t},b_1,b_2),(\bar{t},b_2,b_1)\in {X'}^{\struct{C}}$ whenever $\bar{t}\in X^{\struct{A}'}$.
By the definition of $\Phi'_1$, the quantifier-free part of $\Phi'_1$ is trivially true on $\struct{C}$  for all assignments to the free variables whose range contains both $b_1$ and $b_2$.
Also, by the second line of the definition of $\Phi'_2$, the quantifier-free part of $\Phi'_2$ is trivially true on $\struct{C}$ for all assignments 
to the free variables whose range contains both $b_1$ and $b_2$ but it is not the case that $y_1$ is set equal $b_1$ and $y_2$ is set equal $b_2$ or vice versa.

It is now easy to check that $\struct{C}\models \Phi$ and that the inclusion maps from $\struct{B}_1$ and $\struct{B}_2$ to $\struct{C}$ are embeddings. Hence $\fm(\Phi)$ has the SAP.
\end{proof}
We continue with Theorem~\ref{thm:SAP_lemma}, restated below. 
\saplemma*
\begin{proof} The idea is to replace the default equality predicate with a binary predicate $E$ which interprets as a relational congruence in every model. 
This allows us to simulate the identification of two elements $x$ and $y$ by instead including the pair $(x,y)$ in the relation interpreting $E$, making every amalgam strong.
Now, if there exists a Fra\"{i}ss\'{e} limit $\struct{F}$ for the class $\fm(\Phi)$, then the Fra\"{i}ss\'{e} limit for $\fm(\Gamma(\Phi))$ takes the form of the \emph{relational wreath product} of a pure set with $\struct{F}$, sometimes denoted $\omega \wr \struct{F}$ (see~\cite{bodor2022CSP,pinsker2025three,scow2021ramsey, clemente2015reachability} for more information about this construction).
It is easy to show that the step from $\struct{F}$ to $\omega \wr \struct{F}$ results in the loss of algebraicity for the Fra\"{i}ss\'{e} limit, which translates to the gain of the SAP for the age (\cite[Theorem~4.3.5]{Book}).
We must, however, verify the SAP for $\fm(\Gamma(\Phi))$ by hand since in general there there might not be any structure whose age would equal $\fm(\Phi)$.
Below we give a formal definition of $\Gamma(\Phi)$ and verify the two items in the theorem.

Let $\phi$ be the quantifier-free part of $\Phi$, and let $\sigma$ be its signature. For two tuples $\bar{x}_1,\bar{x}_2$ of the same arity $k$, we use $E(\bar{x}_1,\bar{x}_2)$ as a shortcut for the formula stating that, for every $i\in [k]$, the $i$-th entry of $\bar{x}_1$ and the $i$-th entry of $\bar{x}_2$ are $E$-equivalent. We define $\Gamma(\Phi)$ as 
\begin{align*}
     \forall x,y,z  \big( E(x,x) \wedge    \big(E(x,y) \Rightarrow E(y,x)\big) \wedge  \big(E(x,y) \wedge E(y,z) \Rightarrow E(x,z) \big)\big)
  \\
  {}  \wedge \bigwedge\nolimits_{X\in \sigma}   \forall \bar{x}_1,\bar{x}_2   \big( E(\bar{x}_1,\bar{x}_2) \Rightarrow \big(X(\bar{x}_1)  \Leftrightarrow  X(\bar{x}_2)  \big) 
 \big)  \\
 {}  \wedge  \forall \bar{x}  \big(   \phi(\bar{x})\vee \bigvee\nolimits_{x_1,x_2\in \bar{x}} \big(E(x_1,x_2) \wedge (x_1\neq x_2)\big)    \big).  
\end{align*}  
 
First, we verify item~\eqref{item:sap1}.

``$\Leftarrow$'' Suppose that $\fm(\Gamma(\Phi))$ has the SAP.
Let $\struct{B}_1,\struct{B}_2$ be a one-point amalgamation diagram for $\fm(\Phi)$.
We convert it to a one-point amalgamation diagram $\struct{B}'_1,\struct{B}'_2$ for $\fm(\Gamma(\Phi))$ by letting $E$ interpret as the diagonal relation.
Since $\fm(\Phi)$ has the SAP, there exists a strong amalgam $\struct{C}'$ for $\struct{B}'_1,\struct{B}'_2$ in $\fm(\Gamma(\Phi))$.
By the first two lines in the definition of $\Gamma(\Phi)$, we have that $E^{\struct{C}'}$ is a relational congruence on $\struct{C}'$.
Hence, if $(b_1,b_2)\in E^{\struct{C}'}$ holds for the two elements $b_1\in B_1\setminus B_2$ and $b_2\in B_2\setminus B_1$, then $\struct{C}'/E^{\struct{C}'}$ is isomorphic to both $\struct{B}'_1$ and $\struct{B}'_2$.
In that case an amalgam $\struct{C}$ for $\struct{B}_1,\struct{B}_2$ in $\fm(\Phi)$ can be obtained by gluing $b_1$ and $b_2$.
Otherwise a strong amalgam $\struct{C}$ for $\struct{B}_1,\struct{B}_2$ in $\fm(\Phi)$ can be obtained by taking the $\sigma$-reduct of $\struct{C}'$.

``$\Rightarrow$'' Suppose that $\fm(\Phi)$ has the AP.
Let $\struct{B}_1,\struct{B}_2$ be a one-point amalgamation diagram for $\fm(\Gamma(\Phi))$.
Suppose that there exists $b\in B_1\cap B_2$ such that $(b,b_1)\in E^{\struct{B}_1}$ or $(b,b_2)\in E^{\struct{B}_2}$. 
Then we obtain an amalgam $\struct{C}$ by adding tuples to the relations of $\struct{B}_1\cup \struct{B}_2$ so that $E^{\struct{C}}$ is a relational congruence on $\struct{C}$.
In this case, it is easy to see that $\struct{C}\models \Gamma(\Phi)$.
Otherwise, let $\struct{B}'_1\coloneqq \struct{B}_1/E^{\struct{B}_1}$ and $\struct{B}'_2\coloneqq \struct{B}_2/E^{\struct{B}_2}$.
Since $E^{\struct{B}_1}$ and $E^{\struct{B}_2}$ are relational congruences on $\struct{B}_1$ and $\struct{B}_2$, respectively, we have that $\struct{B}'_1,\struct{B}'_2$ is a one-point amalgamation diagram for $\fm(\Phi)$.
Since $\fm(\Phi)$ has the AP, there exists an amalgam $\struct{C}'$ for $\struct{B}'_1,\struct{B}'_2$ in $\fm(\Phi)$.
Without loss of generality, $C'\subseteq B'_1\cup B'_2$.

If $C'\neq B'_1\cup B'_2$, then $\struct{C}'$ was obtained by gluing $[b_1]_{E^{\struct{B}_1}}$ and $[b_2]_{E^{\struct{B}_2}}$ in $\struct{B}'_1\cup \struct{B}'_2$.
Then we obtain $\struct{C}$ from $\struct{B}_1\cup \struct{B}_2$ by adding $(b_1,b_2)$ and $(b_2,b_1)$ to the relation interpreting $E$.
Since $(b,b_1) \notin E^{\struct{B}_1}$ ans $(b,b_2)\notin E^{\struct{B}_2}$ for every $b\in B_1\cap B_2$, $E^{\struct{C}}$ is a relational congruence on $\struct{C}$.
Now it is easy to see that $\struct{C}$ is a strong amalgam for $\struct{B}_1,\struct{B}_2$ in $\fm(\Gamma(\Phi))$.

Otherwise $C' = B'_1\cup B'_2$.
Then we obtain $\struct{C}$ from $\struct{B}_1\cup \struct{B}_2$ by adding tuples to relations so that, for every $X\in \sigma$, we have $\bar{t}\in X^{\struct{C}}$ if and only if $[\bar{t}]_{E^{\struct{B}_1\cup \struct{B}_2}}\in X^{\struct{C}'}$.
Again it is easy to see that $\struct{C}$ is a strong amalgam for $\struct{B}_1,\struct{B}_2$ in $\fm(\Gamma(\Phi))$.

Second, we verify item~\eqref{item:sap2}. Denote the signatures of $\Phi_1$ and $\Phi_2$ by $\sigma_1$ and $\sigma_2$.

``$\Rightarrow$'' Suppose that $\fm(\Phi_1)|^{\tau}\subseteq \fm(\Phi_2)|^{\tau}$. Let $\struct{A}$ be an arbitrary structure from $\fm(\Gamma(\Phi_1))|^{\tau\cup \{E\}}$, and let $\struct{A}_1$ be an arbitrary structure in $\fm(\Gamma(\Phi_1))$ whose $(\tau\cup \{E\})$-reduct equals $\struct{A}$.
We set $\struct{A}'_1\coloneqq \struct{A}_1/E^{\struct{A}_1}$.
Clearly $\struct{A}'_1\models \Phi_1$.
Let $\struct{A}'$ be the $\tau$-reduct of $\struct{A}'_1$.
Since $\fm(\Phi_1)|^{\tau}\subseteq \fm(\Phi_2)|^{\tau}$, there exists a $\sigma_2$-expansion $\struct{A}'_2$ of $\struct{A}'$ such that $\struct{A}'_2\models \Phi_2$.
We define the $(\sigma_2\cup \{E\})$-expansion $\struct{A}_2$ of $\struct{A}$ so that, for every $X\in \sigma_2$, we have $\bar{t}\in X^{\struct{A}_2}$ if and only if $[\bar{t}]_{E^{\struct{A}_1}}\in X^{\struct{A}_2'}$.
It is easy to see that $\struct{A}_2\models \Gamma(\Phi_2)$, and hence $\struct{A}\in \fm(\Gamma(\Phi_2))|^{\tau\cup \{E\}}$.

``$\Leftarrow$'' Suppose that $\fm(\Gamma(\Phi_1))|^{\tau\cup \{E\}}\subseteq \fm(\Gamma(\Phi_2))|^{\tau\cup \{E\}}$. Let $\struct{A}\in \fm(\Phi_1)|^{\tau}$ be arbitrary, and let $\struct{A}_1$ be an arbitrary structure in $\fm(\Phi_1)$ whose $\tau$-reduct equals $\struct{A}$.
We define $\struct{A}'_1$ as the $(\sigma_1\cup \{E\})$-expansion of $\struct{A}_1$ where $E$ interprets as the diagonal relation.
Clearly $\struct{A}'_1\models \Gamma(\Phi_1)$.
Let $\struct{A}'$ be the $(\tau\cup \{E\})$-reduct of $\struct{A}'_1$.
Since $\fm(\Gamma(\Phi_1))|^{\tau\cup \{E\}}\subseteq \fm(\Gamma(\Phi_2))|^{\tau\cup \{E\}}$, there exists a $(\sigma_2\cup \{E\})$-expansion $\struct{A}'_2$ of $\struct{A}'$ such that $\struct{A}'_2\models \Gamma(\Phi_2)$.
We define $\struct{A}_2$ as the $\sigma_2$-reduct of $\struct{A}'_2$.
It is easy to see that $\struct{A}_2\models  \Phi_2$, and hence $\struct{A}\in \fm(\Phi_2)|^{\tau}$. 
\end{proof}

\begin{example} \label{ex:illuistr}
Let $\struct{G}$ be the disjoint union of countably many copies of the complete graph on two vertices. 
It is easy to see that $\struct{G}$ is homogeneous and that $\Age(\struct{G})=\fm(\Phi)$ for a universal sentence $\Phi$ stating that each model is an undirected simple graph which does not embed any connected graph on three vertices. 
Note that $\Age(\struct{G})$ does not have the SAP.
This can be easily seen by considering the amalgamation diagram $(\struct{B}_1,\struct{B}_2)$ consisting of two edges overlapping on a single vertex.
The only way to obtain an amalgam for $(\struct{B}_1,\struct{B}_2)$ within $\fm(\Phi)$ is to identify $b_1$ and $b_2$.
Within $\fm(\Gamma(\Phi))$, there is the second option of adding the pairs $(b_1,b_2),(b_2,b_1)$ to $E^{\struct{B}_1\cup \struct{B}_2}$, which yields a strong amalgam. 
\end{example}
Finally, we prove Theorem~\ref{thm:inside_out_2}, which is restated below.
\secondinsideout*
\begin{proof} 
The idea is to essentially reverse the proof of Theorem~\ref{thm:inside_out}; we split $\Phi$ into two sentences $\Phi_1$, $\Phi_2$ such that computing amalgams for amalgamation diagrams within $\fm(\Phi)$ translates to verifying instances of the containment $ \fm(\Phi_1)|^{\tau}  \subseteq \fm(\Phi_2)|^{\tau}$. 
To this end, we employ a certain padding technique, which is perhaps more intuitive than the one used in Theorem~\ref{thm:inside_out}.
Every $\tau$-structure $\struct{A}$ will have two (possibly overlapping) parts marked by $L$ and $R$; they correspond to the left and the right part of an amalgamation diagram $\struct{B}_1,\struct{B}_2\in \fm(\Phi) $.
Then $\struct{A} \in  \fm(\Phi_1)|^{\tau} $ holds if and only if $\struct{A}$ represents a valid amalgamation diagram in $\fm(\Phi)$, and $\struct{A} \in \fm(\Phi_2)|^{\tau}$ holds if and only if the said amalgamation diagram has an amalgam $\struct{C}$ in $\fm(\Phi)$. 
Here the signature of $\Phi_1$ will already be $\tau$, so taking $\tau$-reducts of $\struct{A} \in  \fm(\Phi_1)$ is in fact not necessary.
Below we give the details of the construction and verify the equivalence of the four items in the theorem.

Let $\phi(\bar{x})$ be the quantifier-free part of $\Phi$.
We first define the signatures $\tau$ and $\sigma$.
The signature $\tau$ consists of all symbols from $\rho$ together with two fresh unary symbols $L$ and $R$.
The signature $\sigma$ contains all  symbols from $\tau$ and additionally, for every $k$-ary symbol $X\in \rho$, a fresh $k$-ary symbol $X'$.
Next, we define the two sentences $\Phi_1$ and $\Phi_2$.
We write $\text{FREE}(\bar{x})$ as a shortcut for the formula 
    $$\bigvee\nolimits_{x_1,x_2\in \bar{x}} \big(\neg L(x_1)\wedge \neg R(x_2)\big).$$ 
Intuitively, this formula marks all tuples which are not properly contained in the left or the right part of the amalgamation diagram and can thus freely be added to the relations of a potential amalgam; we will call such tuples \emph{free}.
We set:
\begin{align*}
    \Phi_1\coloneqq & \ \forall x \big(L(x) \vee R(x) \big) \wedge  \forall \bar{x} \big(\phi(\bar{x}) \vee \text{FREE}(\bar{x}) \big) \wedge \bigwedge\nolimits_{X\in \rho} \forall \bar{y}   \big(\text{FREE}(\bar{y}) \Rightarrow \neg X(\bar{y}) \big)         , \ \\  
    \Phi_2 \coloneqq & \  \forall x \big(L(x) \vee R(x) \big)  \wedge \forall \bar{x} \ldotp \phi'(\bar{x}) \wedge \bigwedge\nolimits_{X\in \rho} \forall \bar{y} \big(\big(\text{FREE}(\bar{y}) \wedge \neg X(\bar{y}) \big) \vee \big(X(\bar{y})\Leftrightarrow X'(\bar{y}) \big) \big)   ,   
\end{align*}  
where $\phi'$ is obtained from $\phi$ by replacing each atomic $\rho$-formula $X(\bar{x})$ with $X'(\bar{x})$.

``$\eqref{item:in_out_1}\Rightarrow\eqref{item:in_out_4}$''  
Let $\struct{A}\in \fm(\Phi_1) $ be arbitrary, and let $\struct{A}'$ be the $\rho$-reduct of $\struct{A}$.
We obtain the amalgamation diagram $\struct{B}_1,\struct{B}_2$ for $\fm(\Phi)$ by defining $\struct{B}_1$ and $\struct{B}_2$ as the substructures of $\struct{A}'$ on $L^{\struct{A}}$ and $R^{\struct{A}}$, respectively.
Since $\fm(\Phi)$ has the SAP, there exists a strong amalgam $\struct{C}$ for $\struct{B}_1$ and $\struct{B}_2$ in $\fm(\Phi)$.
We define the $\sigma$-expansion $\struct{A}''$ of $\struct{A}$ by setting 
$X^{\struct{A}''}\coloneqq {X'}^{\struct{C}}$ for every $X\in \rho$.
Since $\Phi_1$ prevents free tuples from being in $\rho$-relations while simultaneously imposing $\phi$ on all other tuples,
we clearly have $\struct{A}''\models \Phi_2$, and hence $\struct{A}\in \fm(\Phi_2)|^{\tau}$.

``$\eqref{item:in_out_4}\Rightarrow\eqref{item:in_out_1}$''  
Let $\struct{B}_1,\struct{B}_2$ be an amalgamation diagram for $\fm(\Phi)$.
We obtain the $\tau$-structure $\struct{A}$ from $\struct{B}_1\cup \struct{B}_2$ by setting $L^{\struct{A}}\coloneqq B_1$ and $R^{\struct{A}}\coloneqq B_2$.
Since $\fm(\Phi_1) \subseteq \fm(\Phi_2)|^{\tau}$, there exists a $\sigma$-expansion $\struct{A}'$ of $\struct{A}$ satisfying $\Phi_2$.
Similarly as in the previous case, we obtain a strong amalgam $\struct{C}$ for $\struct{B}_1$ and $\struct{B}_2$ in $\fm(\Phi)$ from $\struct{B}_1\cup \struct{B}_2$ by setting $X^{\struct{C}}\coloneqq {X'}^{\struct{A}'}$ for every $X\in \rho$.

``$\eqref{item:in_out_2}\Rightarrow\eqref{item:in_out_1}$'' Let $\struct{B}_1,\struct{B}_2$ be an amalgamation diagram for $\fm(\Phi)$. Consider the $\tau$-expansions $\struct{B}'_1,\struct{B}'_2$ where $L$ and $R$ both interpret as the full unary relation.
Clearly, $\struct{B}'_1,\struct{B}'_2$ is an amalgamation diagram for $\fm(\Phi_1)$.
Since  $\fm(\Phi_1)$ has the SAP, there exists a strong amalgam $\struct{C}'$ for $\struct{B}'_1$ and $\struct{B}'_2$ in $\fm(\Phi_1)$.
Since $C'=L^{\struct{C}'}=R^{\struct{C}'}$, the $\rho$-reduct $\struct{C}$ of $\struct{C}'$ is a strong amalgam for $\struct{B}_1$ and $\struct{B}_2$ in $\fm(\Phi)$.

 ``$\eqref{item:in_out_1}\Rightarrow\eqref{item:in_out_2}$''  
Let $\struct{B}_1,\struct{B}_2$ be an amalgamation diagram for $\fm(\Phi_1)$.
We obtain two amalgamation diagrams $\struct{B}'_1,\struct{B}'_2$ and $\struct{B}''_1,\struct{B}''_2$ for $\fm(\Phi)$ by taking the $\rho$-reducts of the substructures of $\struct{B}_1,\struct{B}_2$ on the subsets defined by $L$ and $R$, respectively.
Since $\fm(\Phi)$ has the SAP, there exist strong amalgams $\struct{C}'$ and $\struct{C}''$  for the two amalgamation diagrams above.
Without loss of generality, we have $C'= B'_1\cup B'_2$ and $C''= B''_1\cup B''_2$.
We define $\struct{C}$ as the $\tau$-expansion of $\struct{C}'\cup \struct{C}''$ where  $L^{\struct{C}}=B'_1\cup B'_2$ and $R^{\struct{C}}=B''_1\cup B''_2$.
Since $\Phi_1$ prevents free tuples from being in $\rho$-relations while simultaneously imposing $\phi$ on all other tuples, it follows that $\struct{C}$ is a strong amalgam for $\struct{B}_1,\struct{B}_2$ in $\fm(\Phi_1)$.

 ``$\eqref{item:in_out_3}\Rightarrow\eqref{item:in_out_1}$'' Let $\struct{B}_1,\struct{B}_2$ be an amalgamation diagram for $\fm(\Phi)$. Consider the $\sigma$-expansions $\struct{B}'_1,\struct{B}'_2$ where $L$, $R$ both interpret as the full unary relation and the remaining symbols in $\sigma$ interpret as their counterparts in $\rho$.
Clearly, $\struct{B}'_1,\struct{B}'_2$ is an amalgamation diagram for $\fm(\Phi_2)$.
Since $\fm(\Phi_2)$ has the SAP, there exists a strong amalgam $\struct{C}'$ for $\struct{B}'_1$, $\struct{B}'_2$ in $\fm(\Phi_2)$.
Then the $\rho$-reduct $\struct{C}$ of $\struct{C}'$ is a strong amalgam for $\struct{B}_1$ and $\struct{B}_2$ in $\fm(\Phi)$.

 ``$\eqref{item:in_out_1}\Rightarrow\eqref{item:in_out_3}$''  
Let $\struct{B}_1,\struct{B}_2$ be an amalgamation diagram for $\fm(\Phi_2)$.
Let $\struct{B}'_1,\struct{B}'_2$ be the $\rho$-structures with domains $B'_1,B'_2$, respectively, and relations $
X^{\struct{B}'_i}\coloneqq  {X'}^{\struct{B}_i}$ ($i\in [2]$) for every $X\in \rho$.
By the definition of $\Phi_2$,  $\struct{B}''_1,\struct{B}''_2$ is an amalgamation diagram for $\fm(\Phi)$.
Since $\fm(\Phi)$ has the SAP, there exists a strong amalgam $\struct{C}'$ for $\struct{B}'_1$ and $\struct{B}'_2$ in $\fm(\Phi)$.
We obtain a strong amalgam $\struct{C}$ for $\struct{B}_1$ and $\struct{B}_2$ in $\fm(\Phi_2)$ by lifting the $\rho$-relations of $\struct{C}'$ to their $\sigma$-counterparts on all free tuples. 
\end{proof}

\section{Hardness of the Freely Amalgamated RCP} \label{section:cont_hard}

The present section is fully devoted to the proof of Theorem~\ref{thm:containment_hard}, restated below.
\containmenthard* 
\subsection{The square tiling problem} The proof of Theorem~\ref{thm:containment_hard} is by a polynomial-time reduction from the 2-exponential version of the square tiling problem. 
We first state the basic version of this problem, and then its 2-exponential variant.
Consider the signature \emph{$\tau_{\#}$ of grids} consisting of the two binary symbols $\widehat{S}_{1}$ and $\widehat{S}_{2}$. 
For every positive integer $n$, we define the $\tau_{\#}$-structure $\widehat{\struct{Q}}_{n}$ as follows: its domain is $[n]^{2}$, and the relations are 
\begin{align*} 
    \widehat{S}_{1}^{\widehat{\struct{Q}}_{n}} \coloneqq \ & \{((i,j),(i+1,j))\mid i\in [n-1], j\in [n] \}, \\
    \widehat{S}_{2}^{\widehat{\struct{Q}}_{n}} \coloneqq \ & \{((i,j),(i,j+1))\mid i\in [n], j\in [n-1] \}.
\end{align*}  

 \medskip 
\noindent  \underline{\textbf{Square Tiling Problem}}   \\
 \setlength{\tabcolsep}{0pt}    \begin{tabular}{ll}  
  INSTANCE: & \,  A natural number $n$, a finite $\tau_{\#}$-structure $\struct{T}$, and $t\in T$. \\ 
      QUESTION: & \,   Does there exist a homomorphism $h\colon \widehat{\struct{Q}}_{n} \rightarrow \struct{T}$ with $h(1,1)=t$?
\end{tabular}  
\medskip 

Intuitively, the  \emph{grid} structure $\widehat{\struct{Q}}_{n}$ represents an $(n\times n)$-grid and the \emph{tiling} structure $\struct{T}$ represents a finite set of tile types together with certain vertical and horizontal compatibility constraints.
The existence of a homomorphism $h\colon \widehat{\struct{Q}}_{n} \rightarrow \struct{T}$ corresponds to the situation where every point on the grid can be assigned a tile from $T$ while simultaneously satisfying all said compatibility constraints.
The condition $h(1,1)=t$ additionally specifies the tile type for the bottom left corner of the grid. 
Clearly, we could have chosen some more intuitive notation for the symbols $\widehat{S}_{1}$ and $\widehat{S}_{2}$ given their natural meaning, e.g., $V$ and $H$; however, the present notation allows for a more elegant formulation of the proof of Theorem~\ref{thm:containment_hard}. 

The square tiling problem is well-known to be \NP-complete~\cite[Theorem~7.2.1]{lewis1998elements}; the membership in \NP\ is obvious, and the \NP-hardness can be shown by translating the space-time diagrams of polynomial-time Turing machine computations into tilings of grid structures admissible with respect to a fixed tiling structure determined by the Turing machine.
In this reduction, an initial segment of the bottom row specifies the initial configuration, i.e., needs to be a part of the input.
However, the \NP-hardness holds already when the bottom left corner is specified (cf.~\cite[Problem~7.2.2]{lewis1998elements}), which allows us to use the version presented above. 
 
One can further increase the complexity by allowing a succinct encoding of the square.
The input remains the same but now we ask for an admissible tiling of a much larger square, e.g., with $2^{2^n}$ rows and columns. 
In our notation, this would correspond to a homomorphism $h\colon \widehat{\struct{Q}}_{2^{2^n}} \rightarrow \struct{T}$.
It is more or less folklore that, analogously to the natural complete problems based on Turing machines, this yields a decision problem which is complete for the complexity class~\TWONEXPTIME~(cf.~\cite[Section~3.2]{johnson1990catalog}).
We remark that the same problem was used in~\cite{bourhis2016containment} to prove \TWONEXPTIME-hardness of the containment problem for the logic MMSNP. 

\medskip 
\noindent\parbox{\textwidth}{ 
\medskip   
\noindent  \underline{\textbf{2-Exponential Square Tiling Problem}} \\
 \setlength{\tabcolsep}{0pt}    \begin{tabular}{ll}  
  INSTANCE: & \,  A natural number $n$, a finite $\tau_{\#}$-structure $\struct{T}$, and $t\in T$. \\ 
      QUESTION: & \,   Does there exist a homomorphism $h\colon \widehat{\struct{Q}}_{2^{2^n}} \rightarrow \struct{T}$ with $h(1,1)=t$?
\end{tabular}  
\medskip }

\subsection{Proof of Theorem~\ref{thm:containment_hard}}
\begin{proof}[Proof of Theorem~\ref{thm:containment_hard}] 
The idea of the proof is the following. We construct two universal first-order sentences $\Phi_1$ and $\Phi_2$ over the signatures $\sigma_1$ and $\sigma_2$ sharing a common subset $\tau\subseteq \sigma_1\cap \sigma_2$; this subset will correspond to  the signature $\tau_{\#}$ of grids.
Intuitively, the sentences $\Phi_1$ and $\Phi_2$ will have the property that $\fm(\Phi_1)|^{\tau}$ consists of finite unions of copies of substructures of $\widehat{\struct{Q}}_{2^{2^n}}$ and $\fm(\Phi_1)|^{\tau}$ consists of structures which admit a tiling admissible with respect to $\struct{T}$ (despite potentially not being grid-like).  
It immediately follows that  $\fm(\Phi_1)|^{\tau} \subseteq \fm(\Phi_2)|^{\tau}$ if and only if there exists a homomorphism $h\colon \widehat{\struct{Q}}_{2^{2^n}} \rightarrow \struct{T}$. 
Now, recall the definition of completeness for clauses from Section~\ref{section:prelims_logic}.
The crucial part of our construction is that the quantifier-free parts of $\Phi_1$ and $\Phi_2$ are conjunctions of implications each of which is equivalent to a conjunction of (exponentially many) complete clauses. 
This means that both $\fm(\Phi_1)$ and $\fm(\Phi_2)$ are preserved under taking unions, and hence have the FAP.

The signature $\tau$ consists of the binary symbol $O$, the two $4$-ary symbols $S_1,S_2$, and the two $(n+3)$-ary symbols $C_1,C_2$.
The symbol $O$ represents the initial pair $(1,1)$, and the symbols $S_1,S_2$ represent vertical and horizontal successor relations, similarly as $\widehat{S}_{1},\widehat{S}_{2}$.
The symbols $C_1,C_2$ are used to encode the nodes of the $2^{2^n}\times 2^{2^n}$-square grid.
We reveal in advance that the nodes are encoded on pairs of first-order variables, but the encoding is not coordinate-wise; the first (second) entry in a pair of first-order variables does not specify the first (second) coordinate of a grid node.
Instead, the first and the second entry in a pair of first-order variables serve as two binary counters in a representation of the first and the second coordinate of a grid node using atomic $\{C_1,C_2\}$-formulas.
More specifically, we use the symbols $C_1,C_2$ to encode functions $f\colon \{x_1,x_2\}^{n} \rightarrow \{x_1,x_{2}\}$ indexed by pairs $(x_1,x_2)$; an atom $C_k(x_1,x_2,b_1,\dots, b_{n},b)$ with $b,b_1,\dots, b_n\in \{x_1,x_2\}$ is to be interpreted as ``the function value at $(b_1,\dots, b_{n})$ is $b$.''
Note that there are $2^{2^{n}}$-many such functions.

The universal first-order sentence $\Phi_1$ verifies the $2^{2^{n}}$-grid.
In addition to all symbols from $\tau$, its signature $\sigma_1$ contains $2n$-many symbols $\{B_{k}^{i} \mid i\in [n], k\in [2] \}$ of arity $6$ and the six symbols $\{I_{k},U_{k},D_{k} \mid k\in [2] \}$ of arity $2(n+2)$.
The purpose of these symbols is explained below.
We use
$\mathrm{SUCC}(x_1,x_2,b_1,\dots, b_n ,c_1,\dots, c_n )$ as a shortcut for the formula
\begin{align*}  \big(\bigvee\nolimits_{j\in [n]} \bigwedge\nolimits_{i\in [j-1]} \big(
b_i= x_1 \wedge c_i= x_1 \vee b_i= x_2 \wedge c_i= x_2 \big)\big)&  \\ {} \wedge \big(b_j = x_1 \wedge c_j= x_2\big)\wedge   \big( \bigwedge\nolimits_{i\in [n]\setminus [j]} \big(b_i= x_2 \wedge c_i= x_1\big) \big) &.
\end{align*}  
The intended meaning is that $(c_1,\dots, c_n)$ is the successor of $(b_1,\dots, b_n)$ when both tuples are interpreted as numbers encoded in binary in terms of equalities with $x_1$ and $x_2$.
Note that the numbers are read from the right to the left; there is an initial synchronized part of the tuples $(c_1,\dots, c_n)$ and $(b_1,\dots, b_n)$ in the entries $1, \dots,j-1$, followed by an upwards flipped bit at $j$, and by a downwards flipped bit (carryover) in the entries $j+1,\dots,n$.

For $k\in [2]$, we include the following (universally quantified) implications as conjuncts in $\Phi_1$.
We start with a conjunct stating that, for both $k\in [2]$, the $\{C_k\}$-atoms on $(x_1,x_2)$ specify a partial function $f\colon \{x_1,x_2\}^n \rightarrow \{x_1,x_2\}$:
\begin{align*} 
C_{k}(x_1,x_2,b_1,\dots, b_{n},x_1)  \wedge {} &    C_{k}(x_1,x_2,b_1,\dots, b_{n},x_2)  \\ {}\wedge  {} &   \bigwedge\nolimits_{i\in [n]} \big(b_i=x_1\vee b_i=x_{2} \big)  \Rightarrow \bot.
\end{align*}  
Next, we include a conjunct ensuring that $O(x_1,x_2)$ encodes the initial pair $(1,1)$:
\[
O(x_1,x_2)\wedge  \bigwedge\nolimits_{i\in [n]} \big(b_i=x_1\vee b_i=x_{2}) \Rightarrow C_{k}(x_1,x_2,b_1,\dots, b_n,x_1).
\]
Before we proceed further, let us restate formally how the encoding of the nodes of the grid should work.
In each pair $(x_1,x_2)$, the first and the second entry serve as the bits zero and one, respectively, and the value in the $k$-th coordinate $(k\in [2])$ of the node $(x_1,x_2)$ is determined by some function $f\colon \{x_1,x_2\}^n \rightarrow \{x_1,x_2\}$, which is implicitly defined on the tuple $(x_1,x_2)$ by some conjunction
\[\bigwedge\nolimits_{b_1,\dots, b_n\in \{x_1,x_2\}} C_k(x_1,x_2, b_1,\dots, b_n,f(b_1,\dots, b_n)).\]
We already have that, on each pair $(x_1,x_2)$, the conjunction of all $\{C_k\}$-atoms specifies a partial function $f\colon \{x_1,x_2\}^n \rightarrow \{x_1,x_2\}$.
The step from a partial function to a proper function is achieved implicitly in what follows below.

In the final part of $\Phi_1$, we ensure that each atom $S_k(x^{1}_1,x^{1}_2,x^{2}_1,x^{2}_2)$ represents a horizontal or vertical successor pair of grid nodes.
This is where having existentially quantified second-order variables becomes essential.
Recall that above we defined the formula $\mathrm{SUCC}$ to simulate a successor predicate for numbers in $[2^{n}]$ encoded in binary as $n$-tuples over $\{x_1,x_2\}$.
We take a similar approach in simulating a successor predicate for numbers in $[2^{2^n}]$, but this time the numbers are encoded by $2^{n}$-tuples of $2(n+1)$-ary atomic formulas satisfied by the quadruple $(x^{1}_1,x^{1}_2,x^{2}_1,x^{2}_2)$.

We use the symbol $I_k$ to mark the initial (synchronized) part of a pair of $2^{n}$-tuples, i.e., where the function values in terms of $C_k$-atoms coincide on both pairs $(x^{1}_1,x^{1}_2)$ and $(x^{2}_1,x^{2}_2)$.
An atom $I_{k}(x^{1}_1,x^{1}_2,b^{1}_1,\dots, b^{1}_n , x^{2}_1,x^{2}_2, b^{2}_1,\dots, b^{2}_n)$ for $b^{1}_1,\dots, b^{1}_n \in \{x^{1}_1,x^{1}_2\}$ and $b^{2}_1,\dots, b^{2}_n \in \{x^{2}_1,x^{2}_2 \}$ is to be interpreted as ``the function values of $(x^{1}_1,x^{1}_2)$ and 
$(x^{2}_1,x^{2}_2)$ at the arguments specified by $b^{1}_1,\dots, b^{1}_n$ and  $b^{2}_1,\dots, b^{2}_n$, respectively, are identical.''
Furthermore, we use the symbol $U_k$ for the upwards flipped bit, and the symbol $D_k$ for the downwards flipped bit (carryover).
First, for both $k\in [2]$, we include two conjuncts which together force the choice of some upwards flipped bit while keeping the second coordinate ($3-k$) fully synchronized:
\begin{align*}
    S_k(x^{1}_1,x^{1}_2,x^{2}_1,x^{2}_2) \Rightarrow I_{3-k}(x^{1}_1,x^{1}_2,x^{1}_2,\dots, x^{1}_2 , x^{2}_1,x^{2}_2, x^{2}_2,\dots, x^{2}_2 )  \\ {} \wedge \bigwedge\nolimits_{i\in [n]} B_{k}^{i}(x^{1}_1,x^{1}_2,x^{1}_1, x^{2}_1,x^{2}_2,x^{2}_1) \vee B_{k}^{i}( x^{1}_1,x^{1}_2,x^{1}_2, x^{2}_1,x^{2}_2, x^{2}_2),
\end{align*}  
\begin{align*} \bigwedge\nolimits_{i\in [n]} \big(b^{1}_{i}=x^{1}_1 \wedge b^{2}_{i}=x^{2}_1   \vee b^{1}_{i}=x^{1}_2 \wedge b^{2}_{i}=x^{2}_2\big) \wedge   B_{k}^{i}( x^{1}_1,x^{1}_2,b^{1}_i, x^{2}_1,x^{2}_2, b^{2}_i) \\ \Rightarrow U_k(x^{1}_1,x^{1}_2,b^{1}_1,\dots, b^{1}_n , x^{2}_1,x^{2}_2, b^{2}_1,\dots, b^{2}_n ) .
\end{align*} 
Then we include two conjuncts for the backward propagation of the synchronized part and the forward propagation of the downwards flipped bit $(k\in [2])$: 
\begin{align*} 
\big(U_k(x^{1}_1,x^{1}_2,b^{1}_1,\dots, b^{1}_n , x^{2}_1,x^{2}_2, b^{2}_1,\dots, b^{2}_n ) \vee I_k(x^{1}_1,x^{1}_2,b^{1}_1,\dots, b^{1}_n , x^{2}_1,x^{2}_2, b^{2}_1,\dots, b^{2}_n ) \big) \\{}  \wedge 
\mathrm{SUCC}(x^{1}_1,x^{1}_2,c^{1}_1,\dots, c^{1}_n, b^{1}_1,\dots, b^{1}_n)
 \wedge 
\mathrm{SUCC}(x^{2}_1,x^{2}_2,c^{2}_1,\dots, c^{2}_n,b^{2}_1,\dots, b^{2}_n)
\\
\Rightarrow     I_k(x^{1}_1,x^{1}_2,c^{1}_1,\dots, c^{1}_n , x^{2}_1,x^{2}_2, c^{2}_1,\dots, c^{2}_n ) ,
\end{align*}
\begin{align*} 
\big(U_k(x^{1}_1,x^{1}_2,b^{1}_1,\dots, b^{1}_n , x^{2}_1,x^{2}_2, b^{2}_1,\dots, b^{2}_n ) \vee D_k(x^{1}_1,x^{1}_2,b^{1}_1,\dots, b^{1}_n , x^{2}_1,x^{2}_2, b^{2}_1,\dots, b^{2}_n ) \big) \\ {}  \wedge 
\mathrm{SUCC}(x^{1}_1,x^{1}_2,b^{1}_1,\dots, b^{1}_n,c^{1}_1,\dots, c^{1}_n )
  \wedge 
\mathrm{SUCC}(x^{2}_1,x^{2}_2,b^{2}_1,\dots, b^{2}_n,c^{2}_1,\dots, c^{2}_n ) 
\\
\Rightarrow     D_k(x^{1}_1,x^{1}_2,c^{1}_1,\dots, c^{1}_n , x^{2}_1,x^{2}_2, c^{2}_1,\dots, c^{2}_n ) .
\end{align*}
Finally, the following three conjuncts provide an interpretation of $\{U_k,I_K,D_k\}$-atoms in terms of functions encoded by the $C_k$-atoms $(k\in [2])$: 
\begin{align*}
U_k(x^{1}_1,x^{1}_2,b^{1}_1,\dots, b^{1}_n , x^{2}_1,x^{2}_2, b^{2}_1,\dots, b^{2}_n ) 
\Rightarrow  &  \big( C_{k}(x^{1}_1,x^{1}_2,b^{1}_1,\dots, b^{1}_n,x^{1}_1) \\ {} \wedge {} & C_{k}(x^{2}_1,x^{2}_2, b^{2}_1,\dots, b^{2}_n,x^{2}_2)   \big),
\end{align*} 
\begin{align*}
I_k(x^{1}_1,x^{1}_2,b^{1}_1,\dots, b^{1}_n , x^{2}_1,x^{2}_2, b^{2}_1,\dots, b^{2}_n ) 
\Rightarrow  \big( \big(& C_{k}(x^{1}_1,x^{1}_2,b^{1}_1,\dots, b^{1}_n,x^{1}_1) \\  {}  \wedge  {} & C_{k}(x^{2}_1,x^{2}_2, b^{2}_1,\dots, b^{2}_n,x^{2}_1) \big) \\ {} \vee     \big(  C_{k}(x^{1}_1,x^{1}_2,b^{1}_1,\dots, b^{1}_n,x^{1}_2)   \wedge {} & C_{k}(x^{2}_1,x^{2}_2, b^{2}_1,\dots, b^{2}_n,x^{2}_2) \big)  \big) ,
\end{align*}
\begin{align*}
D_k(x^{1}_1,x^{1}_2,b^{1}_1,\dots, b^{1}_n , x^{2}_1,x^{2}_2, b^{2}_1,\dots, b^{2}_n ) 
\Rightarrow  \big( & C_{k}(x^{1}_1,x^{1}_2,b^{1}_1,\dots, b^{1}_n,x^{1}_2) \\ {} \wedge {} & C_{k}(x^{2}_1,x^{2}_2, b^{2}_1,\dots, b^{2}_n,x^{2}_1)    \big). 
\end{align*}

The universal first-order sentence $\Phi_2$ verifies the tiling.
In addition to $\tau$, its signature $\sigma_2$ contains the binary symbol $T_a$ for every $a\in T$.
For $k\in [2]$, we include the following (universally quantified) implications as conjuncts in $\Phi_2$.
First, we include the  implication $O(x,y)\Rightarrow T_t(x,y)$, which ensures that the initial pair $(1,1)$ is assigned the distinguished tile $t\in T$. 
Next, we include conjuncts ensuring the horizontal and vertical consistency of the tiling $(k\in [2])$:
\begin{align*}
    S_k(x^{1}_1,x^{1}_2,x^{2}_1,x^{2}_2) \Rightarrow  \ & \bigvee\nolimits_{(a,b)\in  \widehat{S}_k^{\struct{T}}} T_a(x^{1}_1,x^{1}_2) \wedge T_b(x^{2}_1,x^{2}_2),
\end{align*} 
 and $T_a(x_1,x_2)\wedge T_b(x_1,x_2) \Rightarrow \bot$ for every pair $a,b\in T$ with $a\neq b$.

It is straightforward to verify that each of the above implications is equivalent to a conjunction of (exponentially many) complete clauses; we must only instantiate all disjunctions of equalities in their premises via a suitable substitution of variables and check that the Gaifman graph is complete.
It remains to verify that our reduction is correct; this follows almost directly from the definitions of $\Phi_1$ and $\Phi_2$.

``$\Leftarrow$'' Let $\struct{Q}_{2^{2^n}}$ be the $\tau$-structure with domain $2^{2^n}\times 2^{2^n}$ such that the $C_k$-atoms satisfied by $(i,j)$ correctly encode its value and the $S_k$-atoms correctly determine the horizontal and vertical successor relations. 
Then, by construction, we have $\struct{Q}_{2^{2^n}}\in \fm(\Phi_1)|^{\tau}$.
If $\struct{Q}_{2^{2^n}}\in \fm(\Phi_2)|^{\tau}$, then any expansion of $\struct{Q}_{2^{2^n}}$ satisfying $\Phi_2$ can be used to define a homomorphism $h\colon \widehat{\struct{Q}}_{2^{2^n}} \rightarrow \struct{T}$ satisfying $h(1,1)=t$. 

``$\Rightarrow$'' If there exists a homomorphism $h\colon \widehat{\struct{Q}}_{2^{2^n}} \rightarrow \struct{T}$ satisfying $h(1,1)=t$, then $\fm(\Phi_1)|^{\tau}\subseteq \fm(\Phi_2)|^{\tau}$ holds because we can add a pair $(x,y)$ in $\struct{A}\in \fm(\Phi_1)|^{\tau}$ to $T_a$ if and only if $(x,y)$ encodes $(i,j)$ in terms of $C_k$-atoms and $h(i,j)=a$.
\end{proof}

 \begin{remark} \label{rk:bounds}
    With a slight modification, the proof of Theorem~\ref{thm:containment_hard} can be used to prove \NEXPTIME-hardness for the case where inputs are specified by sets of bounds instead of universal first-order sentences.
    The idea is as follows.
    
    As stated in the proof of Theorem~\ref{thm:containment_hard}, each conjunct in $\Phi_1$ or $\Phi_2$ has the form of an implication that is equivalent to a conjunction of exponentially many complete clauses. 
    By instantiating the equalities in the premises of the implications and imposing some additional equalities which were not relevant for the original construction, the number of free variables in the said complete clauses can be bounded by four.
    Moreover,  the tiles can be stored in binary, e.g., using a single symbol of arity $\lceil \log_2 |T| \rceil+2$ instead of $|T|$-many symbols of arity $2$.
    
    Now, on an input $(n, \struct{T}, t)$ to the 2-exponential square tiling problem, we compute $\Phi_1$ and $\Phi_2$ as in the proof of Theorem~\ref{thm:containment_hard} for the input $(\lceil \log_2 n \rceil, \struct{T}, t)$ and with the modification described above.
    This way, the number of symbols used to represent the grid and their arities will be logarithmic instead of linear; also there is only a single symbol of logarithmic arity representing the tiles.
    We then define $\mathcal{B}_1$ and $\mathcal{B}_2$ as the sets of all $\sigma_1$- and $\sigma_2$-structures with domain $[i]$ for $i\in [4]$  which do not satisfy $\Phi_1$ and $\Phi_2$, respectively.
    By applying basic logarithm laws, we conclude that the sizes of $\mathcal{B}_1$ and $\mathcal{B}_2$ are polynomial in the size of the input to the tiling problem.
    Due to the logarithmic offset of the input parameter $n$, the resulting set of bounds provides a polynomial-time reduction from the 1-exponential square tiling problem (i.e., where $\widehat{\struct{Q}}_{2^{n}}$ is to be tiled instead of $\widehat{\struct{Q}}_{2^{2^n}}$); this problem is \NEXPTIME-complete.
    The above idea also works in combination with Theorem~\ref{thm:inside_out}, because the mapping $\Omega$ only changes the number of symbols and their arities by a constant amount, but the maximal size of a bound increases to $6$. 
\end{remark}
\section{Conditional Decidability of the ADP~\label{section:ramsey}}
The  present section is fully devoted to the proof of Theorem~\ref{theorem:ramsey}, restated below. 
\ramsey*
\subsection{Abstract recolorings} 
 
In the proof of Theorem~\ref{theorem:ramsey}, we will use the notion of an abstract recoloring for universal sentences defined below.
We remark that this definition basically coincides with the one for SNP sentences given in~\cite{barsukov2025containment} and is compatible with the definition of a recoloring for the logic MMSNP given in~\cite{madelaine2010containment,bodirsky2021proof}.

Let $\Phi_1$ and $\Phi_2$ be two universal first-order sentences over signatures $\sigma_1$ and $\sigma_2$ and let $n$ be the largest arity of a symbol in $\sigma_1\cup \sigma_2$.
For $i\in [2]$, denote by $\ncolors(\Phi_i)$ the set of all structures in $\fm(\Phi_i)$ with domain $[m]$ for some $m\leq n$.
For $\tau\subseteq \sigma_1\cap \sigma_2$, a $\tau$-\emph{recoloring} from $\Phi_1$ to $\Phi_2$ is a mapping $\xi$ from $\ncolors(\Phi_1)$ to $\ncolors(\Phi_2)$ with the following two properties.
 First, the $\tau$-reducts of every $\struct{C} \in \ncolors(\Phi_1)$ and its $\xi$-image are identical.
 Secondly, for every $\struct A\in \fm(\Phi_1)$, there is a structure $\xi'(\struct A)\in \fm(\Phi_2)$ on the same domain such that for all $\struct{C} \in \ncolors(\Phi_1)$ and every  $e\in \binom{\struct{A}}{\struct{C}}$ we have $e\in \binom{\xi'(\struct{A})}{\xi(\struct{C})}$. 
 In other words, the following  extension $\xi'$ of $\xi$ is a well-defined mapping from $\fm(\Phi_1)$ to $\fm(\Phi_2)$: 
 \begin{center} \vspace{0.5em}
    {\it \parbox{0.7\textwidth}{For every $\struct{A}\in \fm(\Phi_1)$, the structure $\xi'(\struct{A})$ on the same domain as $\struct{A}$ is obtained by replacing for every  $\struct{C} \in \ncolors(\Phi_1)$ and every embedding $e\in \binom{\struct{A}}{\struct{C}}$ the substructure $e(\struct{C})$ of $\struct A$ by $e(\xi(\struct{C}))$.}  
    } \vspace{0.5em}
\end{center} 
Note that the second formulation makes it clear that $\xi'(\struct A)$ is unique, as $n$ is the largest arity of a symbol in $\sigma_1\cup \sigma_2$.
For the same reason, we have that $\struct{A}|^{\tau} = \xi'(\struct A)|^{\tau}$.  
Also observe that in the second  formulation the well-definedness of $\xi'$ is a non-trivial property since the required replacements might be on overlapping substructures. 

\begin{lemma}\label{lemma:estimate}
    The existence of a $\tau$-recoloring from $\Phi_1$ to $\Phi_2$ can be tested non-deterministically in time $\mathcal{O}\bigl(2^{2^{\max(|\Phi_1|,|\Phi_2|)}}\bigr)$, where $|\Phi_i|$ denotes the size of $\Phi_i$.    
\end{lemma}
\begin{proof} For $i\in [2]$, let:
\begin{itemize}
    \item $\sigma_i$ be the signature of $\Phi_i$;
    \item $n_i$ be the maximal arity of a symbol in $\sigma_i$;
    \item ${\ell}_i$ be the number of clauses in $\Phi_i$;
    \item $w_i$ be the maximum number of variables per clause in $\Phi_i$. 
\end{itemize} 
Set $n\coloneqq \max(n_1,n_2)$.
Note that $|\ncolors(\Phi_1)|$ is bounded by the number of $\sigma_1$-structures on at most $n$ elements.
    Therefore, it is at most $c\coloneqq2^{|\sigma_1| n^{n_1}}$.
    Let $\xi$ be any mapping from $\ncolors(\Phi_1)$ to $\ncolors(\Phi_2)$.
    We can verify the recoloring property in three steps as follows. 

    First, we check that, for every $\struct{C}\in\ncolors(\Phi_1)$, the $\tau$-reducts of $\struct{C}$ and $\xi(\struct{C})$ are identical.
    This can be done by inspecting the $\xi'$-image of each element of $\ncolors(\Phi_1)$.
    Since $|\ncolors(\Phi_1)|\leq c$, it can be checked in time $\mathcal{O}\bigl(|\sigma_1| n^n c\bigr)$.
    Suppose now that this condition is satisfied.

    Next, we check that, for every pair $\struct{C}_1,\struct{C}_2\in\ncolors(\Phi_1)$, every partial isomorphism $i$ from $\struct{C}_1$ to $\struct{C}_2$ is also a partial isomorphism from $\xi(\struct{C}_1)$ to $\xi(\struct{C}_2)$.
    To this end, we must go through all pairs of elements of $\ncolors(\Phi_1)$ and through all their substructures. 
    This can be done in time $\mathcal{O}\bigl(c^2 (2n)^{2n}|\sigma_2| \bigr)$.  
    Suppose now that this condition is satisfied as well.

    Finally, we check that the extension $\xi'$ maps $\fm(\Phi_1)$ to $\fm(\Phi_2)$.
    If this is not the case, then $\xi'$ is well-defined but violates a clause of $\Phi_2$ in the image a structure that originally satisfied all clauses of $\Phi_1$.
More specifically, there exists a $\sigma_2$-structure $\struct{A}$ of domain size at most $w_2$ such that $\xi'(\struct{A})\notin \fm(\Phi_2)$ while $\struct{A} \in \fm(\Phi_1)$.
To test whether this happens, we go through all $\sigma_1$-structures $\struct{A}$ with domain $A=[m]$ for some $m\leq w_2$; there are at most $2^{w_2^n \smash{|\sigma_1|}}$ of them.
Then, for each such structure $\struct{A}$, we test in time $\mathcal{O}\bigl(\ell_1 w_2^{w_1}|\sigma_1| w_1^n+\ell_2 w_2^{w_2}|\sigma_2| w_2^n\bigr)$ the satisfiability of $\Phi_1$ and $\Phi_2$ in $\struct{A}$ and in $\xi'(\struct{A})$, respectively.
In sum, the total time needed is in $\mathcal{O}\bigl(2^{w_2^n|\sigma_1|}\bigl(\ell_1+\ell_2\bigr)w_2^{w_1+w_2}(w_1+w_2)^n(|\sigma_1|+|\sigma_2|)\bigr)$. 
\end{proof}

\subsection{Proof of Theorem~\ref{theorem:ramsey}} 

\begin{proof}[Proof of Theorem~\ref{theorem:ramsey}]   
Let $g\colon \mathbb{N} \rightarrow \mathbb{N}$ be any function with the property that, if $\Phi$ is a universal first-order sentence over a finite relational signature $\sigma$ with $|\Phi|=n$ and such that $\fm(\Phi)$ has the SAP, then there exists a universal first-order sentence $\Phi^+$ over a (potentially larger) finite relational signature $\sigma^+$ such that: 
\begin{itemize}
    \item $|\Phi^+|\leq g(n)$;
    \item $\fm(\Phi) = \fm(\Phi^+)|^{\sigma}$;
    \item $\fm(\Phi^+)$ has both the RP and the SAP.
\end{itemize}

Recall the functions $\Gamma$ and $\Delta$  from Theorem~\ref{thm:SAP_lemma} and Theorem~\ref{thm:inside_out_2}.
Given a universal first-order sentence $\Phi$ over some finite relational signature, we set  $\Phi_i\coloneqq \Delta_i(\Gamma(\Phi))$ for  $i\in [2]$, and we choose $\tau$ as in Theorem~\ref{thm:inside_out_2}.
Then $\fm(\Phi)$ has the AP if and only if one of the following three equivalent conditions holds:
\begin{itemize}
    \item $\fm(\Phi_1)$ has the SAP;
    \item $\fm(\Phi_2)$ has the SAP;
    \item $\fm(\Phi_1)|^{\tau} \subseteq \fm(\Phi_2)|^{\tau}$.
\end{itemize}
Our conditional decision procedure for the ADP goes as follows. We guess for both $i\in [2]$ a sentence $\Phi_i^+$ of size at most $g(|\Phi_i|)$ such that $\fm(\Phi_i)|^{\tau} = \fm(\Phi_i^+)|^{\tau}$ and there exists a $\tau$-recoloring from $\Phi_1^+$ to $\Phi_2^+$.
By Lemma~\ref{lemma:estimate}, the existence of a $\tau$-recoloring from $\Phi_1^+$ to $\Phi_2^+$ can be tested in non-deterministic 2-exponential time, which leads to the desired upper bound on the runtime of the procedure.

Now, if a $\tau$-recoloring from $\Phi_1^+$ to $\Phi_2^+$ exists, then it is not hard to see that $\fm(\Phi^+_1)|^{\tau} \subseteq \fm(\Phi^+_2)|^{\tau}$ holds, because a recoloring is nothing but a uniform form of reduct containment.
Then it follows immediately that $\fm(\Phi)$ has the AP.
On the other hand, if $\fm(\Phi)$ has the AP, then there must exist suitable $\Phi_1^+$ and $\Phi_2^+$ for which $\fm(\Phi^+_1)|^{\tau} \subseteq \fm(\Phi^+_2)|^{\tau}$ is equivalent to the existence of a $\tau$-recoloring from $\Phi_1^+$ to $\Phi_2^+$.
We will now go through this argument in detail.

``$\Rightarrow$'' Suppose that, for both $i\in [2]$, there exists a sentence $\Phi_i^+$ of size at most $g(|\Phi_i|)$ with $\fm(\Phi_i)|^{\tau} = \fm(\Phi_i^+)|^{\tau}$ and a $\tau$-recoloring $\xi$ from $\Phi_1^+$ to $\Phi_2^+$.
We show that $\fm(\Phi^+_1)|^{\tau} \subseteq \fm(\Phi^+_2)|^{\tau}$.

To this end, let $\struct{A}\in \fm(\Phi^+_1)|^{\tau}$ be arbitrary; then there exists an expansion $\struct{A}'\in \fm(\Phi^+_1)$.
By the definition of a $\tau$-recoloring we have $\xi'(\struct{A}')\in \fm(\Phi_2)$. 
It also follows from this definition that for all $\struct{C} \in \ncolors(\Phi^+_1)$ and every embedding $e\in \binom{\struct{A}}{\struct{C}}$ we have $e\in \binom{\xi'(\struct{A})}{\xi(\struct{C})}$.
By our choice of $n$ in the definition of a $\tau$-recoloring, this implies that the $\tau$-reduct of $\xi'(\struct{A}')$ coincides with $\struct A$. Hence $\xi'(\struct{A}')$ witnesses that  $\struct{A}\in \fm(\Phi^+_2)|^{\tau}$.
Since $\fm(\Phi_i)|^{\tau} = \fm(\Phi_i^+)|^{\tau}$ for both $i\in [2]$, it follows that $\fm(\Phi_1)|^{\tau} \subseteq \fm(\Phi_2)|^{\tau}$.
By Theorem~\ref{thm:SAP_lemma} and Theorem~\ref{thm:inside_out_2}, $\fm(\Phi)$ has the AP.

``$\Leftarrow$'' Suppose that $\fm(\Phi)$ has the AP.
By Theorem~\ref{thm:SAP_lemma} and Theorem~\ref{thm:inside_out_2}, it follows that $\fm(\Phi_i)$ has the SAP for both $i\in [2]$ and that $\fm(\Phi_1)|^{\tau} \subseteq \fm(\Phi_2)|^{\tau}$.
By assumption, for both $i\in [2]$, there exists a sentence $\Phi_i^+$ of size at most $g(|\Phi_i|)$ with $\fm(\Phi_i)|^{\tau} = \fm(\Phi_i^+)|^{\tau}$ and such that $\fm(\Phi_i^+)$ has both the RP and the SAP.
It follows immediately that $\fm(\Phi^+_1)|^{\tau} \subseteq \fm(\Phi^+_2)|^{\tau}$.
We claim that there exists a $\tau$-recoloring from $\Phi_1^+$ to $\Phi_2^+$.

By Theorem~\ref{theorem:fraisse_2}, there exist homogeneous Ramsey structures $\struct{F}_1$ and $\struct{F}_2$ such that $\fm(\Phi^+_1)=\Age(\struct{F}_1)$ and $\fm(\Phi^+_2)=\Age(\struct{F}_2)$.
Then, $\Age(\struct{F}_1|^{\tau}) \subseteq \Age(\struct{F}_2|^{\tau})$.
By the $\omega$-categoricity of $\struct{F}_2|^{\tau}$, Lemma~\ref{lemma:compactness} implies that there exists an embedding $e$ from $\struct{F}_1|^{\tau}$ to $\struct{F}_2|^{\tau}$. 
By Theorem~\ref{th:canonical_ramsey}, we may assume that $e$ is canonical as a function from $\struct{F}_1$ to $\struct{F}_2$ since the former is a homogeneous Ramsey structure.
Let $\xi'$ be the mapping from $\fm(\Phi^+_1)$ to $\fm(\Phi^+_2)$ defined as follows.
Let $n$ be as in the definition of a $\tau$-recoloring.
Given $\struct{A}\in \fm(\Phi^+_1)$, consider an arbitrary embedding $i$ from $\struct{A}$ to $\struct{F}_1$.
Then, we define $\xi'(\struct{A}) \in \fm(\Phi^+_2)$ with domain $A$ whose relations are defined through their preimages under $e\circ i$.   
Let $\xi$ be the restriction of $\xi'$ to $\ncolors(\Phi^+_1)$.
We claim that $\xi'$ arises from $\xi$ as in the definition of a recoloring. 

First, the $\tau$-reduct of any   $\struct{C}\in \ncolors(\Phi_1)$  and its $\xi$-image are identical since $\xi(\struct{C})=\xi'(\struct{C})$ is obtained by pulling back relations via the $\tau$-embedding $e\circ i$.  
For the second part of the definition pertaining to $\xi'$, let $\struct A \in \fm(\Phi^+_1)$, $\struct{C}\in\ncolors(\Phi^+_1)$,  and $f\in \binom{\struct A}{\struct{C}}$
be arbitrary; we claim that $f\in \binom{\xi'(\struct A)}{\xi(\struct{C})}$. By definition, the relations of $\xi'(\struct A)$ and of $\xi(\struct{C})$ are obtained by first embedding $\struct A$ and $\struct{C}$  into $\struct{F}_1$, then applying $e$, and then pulling back the relations from $\struct{F}_2$. This definition does not depend on the choice of the embeddings into $\struct{F}_1$, by the homogeneity of  $\struct{F}_1$ and the canonicity of $e$ --- the claim follows. \end{proof}

\section{Undecidability of Homogenizability~\label{section:undecidability}}
The present section is fully devoted to the proof of Theorem~\ref{thm:undecidability}, restated below. 
\undecidability*
  
The proof of Theorem~\ref{thm:undecidability} is by a polynomial-time reduction from the problem of testing the regularity of context-free languages presented in terms of CFGs.
It is well-known that this problem is undecidable; see, e.g., Theorem~6.6.6 in~\cite{shallit2008second}.  
 
\medskip  
\noindent  \underline{\textbf{Regularity of Context-Free Languages}}   \\
 \noindent\setlength{\tabcolsep}{0pt}\begin{tabular}{ll}  
  INSTANCE: & \,  A CFG $\fancyfont{G}$. \\ 
      QUESTION: & \,   Is $L(\fancyfont{G})$ regular?
\end{tabular}  
\medskip

\subsection{The proof of Theorem~\ref{thm:undecidability}} \label{section:encoding}

We first explain our encoding of context-free grammars and deterministic finite automata into finitely bounded classes (Lemma~\ref{lemma:correspondence1} and Lemma~\ref{lemma:correspondence2}).
We then prove a technical result, Lemma~\ref{lemma:regularity}, from which Theorem~\ref{thm:undecidability} follows immediately due to the fact that the regularity problem for context-free languages is undecidable.

Let $\fancyfont{G}=(N,\Sigma, P,S)$ be an arbitrary CFG. 
We define the signatures $\tau_{\Sigma}$ and $\tau_{N}$ as follows.
The signature $\tau_{\Sigma}$ consists of the unary symbols $I,T$ and the binary symbols $R_a$ for every $a\in \Sigma$, and the signature $\tau_{N}$ consists of a binary symbol $R_a$ for every element $a\in N$. 
For $a_1\dots a_n\in \Sigma^{+}$, we set 
\[\phi_{a_1\dots a_n}(x_1,\dots, x_{n+1})\coloneqq 
\bigwedge\nolimits_{i\in [n]} R_{a_i}(x_i,x_{i+1}).\] 
Now, let $\Phi^{+}_\fancyfont{G}$ be the universal Horn sentence over the signature $\tau_{\Sigma}\cup\tau_{N}$ whose quantifier-free part contains, for every $(A\to w) \in P$, the Horn clause  
   \begin{align} 
 	\phi_{w}(x_1,\dots, x_{|w|+1})  \Rightarrow  R_A(x_1,x_{|w|+1}),  \label{eq:terminal}
  \end{align}  
	and additionally the Horn clause  
         \begin{align} 
		I(x_1)   \wedge R_{S}(x_1,x_{2})  \wedge T(x_{2}) \Rightarrow\bot.  \label{eq:terminal_last}
         \end{align}  
Then, define  $\Phi_\fancyfont{G}$ as the Datalog sentence obtained from $\Phi^{+}_\fancyfont{G}$ by existentially quantifying all symbols from $\tau_{N}$ upfront.
The above encoding of CFGs into Datalog sentences is standard (Exercise~12.26 in \cite{abiteboul1995foundations}), and the correspondence provided by Lemma~\ref{lemma:correspondence1} below can be shown via a straightforward induction; we refer the interested readers to~\cite{schrottenloher2022universal}.
 \begin{lemma} \label{lemma:correspondence1} 
   For every $\tau_{\Sigma}$-structure $\struct{A}$,  the following are equivalent:
 \begin{itemize}
     \item  $\struct{A} \models \Phi_\fancyfont{G}$;
    \item for every $w \in L(\fancyfont{G})$:   $$\struct{A} \models \forall x_{1},\dots ,x_{|w|+1}    \big(   I(x_1)   \wedge   \phi_{w}(x_1,\dots, x_{|w|+1}) \wedge  T(x_{|w|+1}) \Rightarrow \bot \big). $$
 \end{itemize} 
\end{lemma} 

 Next, let $\fancyfont{A}=(Q,\Sigma,\delta,q_0,F)$ be a DFA.
 The signature $\tau_{\Sigma}$ is defined as before, and the signature $\tau_{Q}$ consists of the unary symbols $R_q$ for every $q\in Q$ that is \emph{reachable from $q_0$}, i.e., there exists a word $a_1\dots a_n\in \Sigma^{+}$ such that
 $q=\delta(a_n,\dots\delta(a_{2},\delta(a_{1},q_0))\dots)$. 
 Note that $q_0$ is not necessarily  reachable from itself. 
 Let $\Phi^{+}_{\fancyfont{A}}$ be the universal Horn sentence over the signature $\tau_{\Sigma}\cup \tau_{Q}$ whose quantifier-free part contains: for every $a\in \Sigma$,  the Horn clause  
 \[
 I(x_1)\wedge   R_a(x_1,x_2) \Rightarrow R_{\delta(q_0,a)}(x_2),
 \] 
 for every $(q,a)\in Q\times \Sigma$ such that $q$ is reachable from $q_0$, the Horn clause 
 \[
		R_{q}(x_1)\wedge   R_a(x_1,x_2)  \Rightarrow  R_{\delta(q,a)}(x_2),
\]
and, for every $q\in F$ that is reachable from $q_0$, the Horn clause
 \[
 R_q(x_1) \wedge T(x_1)\Rightarrow \bot.
 \]  
Then, define $\Phi_{\fancyfont{A}}$ as the monadic Datalog sentence obtained from $\Phi^{+}_{\fancyfont{A}}$ by existentially quantifying all symbols from $\tau_{Q}$ upfront.
The following lemma can be proved similarly as Lemma~\ref{lemma:correspondence1}, and its proof is omitted as well.
 
 \begin{lemma} \label{lemma:correspondence2} For every $\tau_{\Sigma}$-structure $\struct{A}$, the following are equivalent
 \begin{itemize}
     \item  $\struct{A} \models \Phi_{\fancyfont{A}}$;
    \item for every $w \in L(\fancyfont{A})$:  
 $$ \struct{A} \models \forall x_{1},\dots ,x_{|w|+1}    \big(   I(x_1)   \wedge   \phi_{w}(x_1,\dots, x_{|w|+1}) \wedge  T(x_{|w|+1}) \Rightarrow \bot \big). $$
 \end{itemize} 
\end{lemma} 

We are now ready to state and prove the main lemma of this section. 
\begin{lemma} \label{lemma:regularity} Let $\fancyfont{G}$ be a context-free grammar. Then the following are equivalent:
\renewcommand{\theenumi}{\alph{enumi}}
\begin{enumerate}  
    \item \label{eq:cfg_item2} $\Phi_\fancyfont{G}$ is equivalent to connected monadic Datalog sentence.
    \item \label{eq:cfg_item4} $\fm(\Phi_\fancyfont{G})$ is the CSP of a finite structure. 
    \item \label{eq:cfg_item9} $\fm(\Phi_\fancyfont{G})$ is the age of a finitely bounded-homogenizable structure.
    \item \label{eq:cfg_item8} $\fm(\Phi_\fancyfont{G})$ is the age of a reduct of a finitely bounded homogeneous Ramsey structure.      
    \item \label{eq:cfg_item3} $\Phi_\fancyfont{G}$ is equivalent to a GMSNP sentence. 
    \item \label{eq:cfg_item7} $\fm(\Phi_\fancyfont{G})$ is the CSP of an $\omega$-categorical structure. 
    \item \label{eq:cfg_item6} $\fm(\Phi_\fancyfont{G})$ is the age of an $\omega$-categorical structure.  
    \item \label{eq:cfg_item1} $L(\fancyfont{G})$ is regular.
\end{enumerate}
\end{lemma}

\begin{remark}
     From every CFG $\fancyfont{G}=(N,\Sigma,P,S)$ one can compute in polynomial time a CFG $\fancyfont{G}'=(N',\Sigma,P',S)$ satisfying $L(\fancyfont{G})=L(\fancyfont{G}')$ and such that, for each  $(A \rightarrow w)\in P'$, the length of $w$ is $\leq 2$.
     This can be done by iteratively introducing new auxiliary non-terminal symbols for each production rule and splitting it into two strictly shorter rules.
     Consequently, Theorem~\ref{thm:undecidability} holds even under the additional restriction to connected Datalog sentences $\Phi$ with at most $3$ variables per clause.
\end{remark}

\begin{proof}[Proof of Lemma~\ref{lemma:regularity}]

``$\eqref{eq:cfg_item2} \Rightarrow \eqref{eq:cfg_item3}$'' This direction is trivial.

``$\eqref{eq:cfg_item4} \Rightarrow \eqref{eq:cfg_item7}$'' This direction is also trivial. 

``$\eqref{eq:cfg_item8} \vee \eqref{eq:cfg_item9}\Rightarrow\eqref{eq:cfg_item6}$'' This direction is well-known and easy to see~\cite{Hodges}.

``\eqref{eq:cfg_item3}$\,\Rightarrow\,$\eqref{eq:cfg_item8}'' 
Since $\Phi_\fancyfont{G}$ is equivalent to a GMSNP sentence, by the first part of Theorem~\ref{theorem:gmsnp_homogenizable}, $\Phi_\fancyfont{G}$ is equivalent to a disjunction $\Phi_1\vee \cdots \vee \Phi_k$ of connected GMSNP sentences.
We may assume that $k$ is minimal with this property.
We claim that $k=1$.
Suppose, on the contrary, that $k\geq 2$.
Recall the definition of connectedness for SNP sentences from Section~\ref{section:prelims_logic}.
Note that $\Phi_\fancyfont{G}$ is connected, and hence $\fm(\Phi_\fancyfont{G})$ is preserved by disjoint unions.
By the minimality of $k$, there exist $\struct{A}_1,\struct{A}_2 \in \fm(\Phi_\fancyfont{G})$ such that 
$\struct{A}_i \models \Phi_i\wedge \bigwedge_{j\in [k]\setminus \{i\}} \neg \Phi_j$ for $i\in \{1,2\}$.
But then the disjoint union of $\struct{A}_1$ and $\struct{A}_2$ does not satisfy $\Phi_\fancyfont{G}$, a contradiction.
Hence, $k=1$.
Now the statement follows from the second part of Theorem~\ref{theorem:gmsnp_homogenizable}.

``$\eqref{eq:cfg_item1}\Rightarrow\eqref{eq:cfg_item2}\wedge\eqref{eq:cfg_item4}\wedge\eqref{eq:cfg_item9}$'' By Theorem~\ref{theorem:myhil_nerode}, there exists a DFA $\fancyfont{A}$ such that $L(\fancyfont{G})=L(\fancyfont{A})$. 
By Lemma~\ref{lemma:correspondence1} and Lemma~\ref{lemma:correspondence2}, we have $\fm(\Phi_\fancyfont{G})= \fm(\Phi_\fancyfont{A}).$
This implies item~\eqref{eq:cfg_item2} because $\Phi_\fancyfont{A}$ is in connected monadic Datalog.
In fact, the sentence falls into an even stricter fragment known as \emph{caterpillar Datalog}~\cite{erdos2013caterpillar}.
By Theorem~4.1 in \cite{erdos2013caterpillar}, there exists a finite structure whose CSP equals $\fm(\Phi_\fancyfont{A})$. This implies item~\eqref{eq:cfg_item4}.
Finally, we show item~\eqref{eq:cfg_item9}.  
 
Recall the definition of completeness for clauses from Section~\ref{section:prelims_logic}.
Note that the universal sentence $\Phi^{+}_{\fancyfont{A}}$ is complete, and hence $\fm(\Phi^{+}_{\fancyfont{A}})$ is closed under unions, and therefore has the FAP.
Let $\struct{F}$ be the Fra\"{i}ss\'{e} limit of $\fm(\Phi^{+}_{\fancyfont{A}})$ (provided by Theorem~\ref{theorem:fraisse_2}).
By the definition of $\fm(\Phi^{+}_{\fancyfont{A}})$, the signature $\tau_{Q}$ only contains unary symbols $R_{q}$ for those $q\in Q$ which are reachable from $q_0$.
Let $Q_0\subseteq Q$ be the set of states reachable from $q_0$ and, for every $q\in Q_0$, let $w_q\in \Sigma^{+}$ be an arbitrary word witnessing that $q$ is reachable from $q_0$.
Note that, if $q_1, q_2\in Q_0$ are distinct, then $w_{q_1}\neq w_{q_2}$.
For every $w\in \Sigma^{+}$, consider the unary formula
\begin{align*}
    \phi^I_w(x_1)\coloneqq\ & \exists x_2,\dots, x_{|w|+1}\big( I(x_{|w|+1})\wedge \phi_{w}(x_{|w|+1}, \dots, x_1) \big).
\end{align*}
We show that, for every $q\in Q_0$, the unary formulas $R_q(x)$ and $\phi^I_{w_q}(x)$ define the same relation in $\struct{F}$.
Then, for every $q\in Q_0$, the relation $R_{q}^{\struct{F}}$ is first-order definable in the $\tau_{\Sigma}$-reduct of $\struct{F}$.  
This implies item~\eqref{eq:cfg_item9}. 

First, suppose that $\struct{F}\models \phi^I_{w_q}(x)$ for some $x\in F$.
Since $w_q$ witnesses that $q$ is reachable from $q_0$ and $\struct{F}\models \Phi^{+}_{\fancyfont{A}}$, it follows that $\struct{F}\models R_q(x)$.
Next, suppose that $\struct{F}\models R_q(x)$ for some $x\in F$.
Since $\fancyfont{A}$ is deterministic, by the definition of $\Phi^{+}_{\fancyfont{A}}$, there exists a structure $\struct{A}$ in the signature of $\struct{F}$ with domain $\{x_{|w_q|+1}, \dots, x_1\}$ that satisfies $\struct{A}\models \Phi^{+}_{\fancyfont{A}}\wedge I(x_{|w_q|+1})\wedge \phi_{w_q}(x_{|w_q|+1}, \dots, x_1)$ and whose substructure on $\{x_1\}$ is isomorphic to the substructure of $\struct{F}$ on $\{x\}$. 
Since $\struct{A}\models \Phi^{+}_{\fancyfont{A}}$, there exists an embedding $e\colon \struct{A} \rightarrow \struct{F}$. Since $\struct{F}$ is homogeneous, there exists an automorphism $f$ of $\struct{F}$ such that $f\circ e(x_1)=x$.
Since automorphisms preserve first-order definable relations, it follows that $\struct{F}\models \phi^I_{w_q}(x)$. 

``$\eqref{eq:cfg_item7}\vee \eqref{eq:cfg_item6}\Rightarrow\eqref{eq:cfg_item1}$'' Let $\struct{D}$ be an $\omega$-categorical structure such that $\fm(\Phi_\fancyfont{G})=\Age(\struct{D})$ or $\fm(\Phi_\fancyfont{G})=\CSP(\struct{D})$. 
Suppose, on the contrary, that $L(\fancyfont{G})$ is not regular.
For every $w\in \Sigma^{+}$, consider the unary formula
\begin{align*} 
\phi^T_w(x_1)\coloneqq\ & \exists x_2,\dots, x_{|w|+1}\big( \phi_{w}(x_1,\dots, x_{|w|+1}) \wedge T(x_{|w|+1}) \big).
\end{align*}
Since $L(\fancyfont{G})$ is not regular, by Theorem~\ref{theorem:myhil_nerode}, the Myhill-Nerode equivalence relation $\sim_{L(\fancyfont{G})}$ has infinitely many classes. 
Let $w_1,\dots$ be the representatives of the classes.
For every $i\geq 1 $ let $R_i\subseteq D$ be the unary relation defined by $\phi^I_{w_i}(x_1)$ in $\struct{D}$.
For every $i\geq 1$, the formula $\phi^I_{w_i}(x_1)$ is satisfiable in $\struct{D}$ because it does not contain any $T$-atom.
Since first-order formulas are preserved under automorphisms, for every $i\geq 1$, $R_i$ is a non-empty union of unary orbits.
By the definition of $\sim_{L(\fancyfont{G})}$, for every pair $i\neq j$, there exists $w \in \Sigma^{*}$ such that $|\{w_i w,w_j w\}\cap L(\fancyfont{G})|=1$. 
By Lemma~\ref{lemma:correspondence1}, exactly one of the relations $R_i$ and $R_j$ contains an element satisfying $\phi^T_{w}(x_1)$ in $\struct{D}$.
In other words, there exists an orbit that is contained in one of the two relations but not in the other.
This can only be the case for all $i\neq j$ if there are infinitely many unary orbits, a contradiction to $\omega$-categoricity. Therefore, $L(\fancyfont{G})$ is regular.  
\end{proof}

%% file: appendix.tex
 \section{Infinite-Domain Constraint Satisfaction Problems}   \label{subs:connections}
The logic class SNP is an expressive fragment of existential second-order logic and thus, by Fagin's Theorem, of the complexity class \textup{\textsf{NP}}. 
Despite the name, SNP already has the full power of \textup{\textsf{NP}}, in the sense that every problem in \textup{\textsf{NP}} is equivalent to a problem in SNP under polynomial-time reductions~\cite{feder1998computational}.
In addition, this logic class has many connections to CSPs. 
The basic link from SNP to CSPs is that every sentence of the monotone fragment of SNP defines a finite disjoint union of CSPs of (possibly infinite) relational structures~\cite{Book}. 
There are, however, some more nuanced connections, such as the one that led to the formulation of the \emph{Feder--Vardi conjecture}, now known as the finite-domain CSP dichotomy theorem~\cite{zhuk2020proof}.

Feder and Vardi~\cite{feder1998computational} showed that the \emph{Monotone Monadic} fragment of SNP (MMSNP) exhibits a dichotomy between \textup{\textsf{P}} and \textup{\textsf{NP}}-completeness if and only if the seemingly less complicated class of all finite-domain CSPs exhibits such a dichotomy, they also conjectured that this is the case.
MMSNP contains all finite-domain CSPs, and many other interesting combinatorial problems, e.g., whether the vertices of a given graph can be vertex-2-colored while omitting any monochromatic triangle~\cite{madelaine2003some}.
Their conjecture was confirmed in 2017 independently by Bulatov and Zhuk~\cite{bulatov2017dichotomy,ZhukFVConjecture}.

There is a yet unconfirmed generalization of the Feder--Vardi conjecture, to the CSPs of reducts of countable finitely bounded homogeneous structures, formulated by Bodirsky and Pinsker in 2012~\cite{bodirsky2021projective}. \footnote{The original arXiv version of the article~\cite{bodirsky2021projective} from~2014 contains the first explicit mention of the Bodirsky--Pinsker conjecture; the peer-reviewed version of~\cite{bodirsky2021projective} appeared much later, in 2021.}
Here we refer to it as the \emph{Bodirsky--Pinsker conjecture}.  
Roughly said, the condition imposed on the structures within the scope of the Bodirsky--Pinsker conjecture ensures that the CSP is in \textup{\textsf{NP}} and that it can be parametrized by a structure enjoying some of the  universal-algebraic properties that have played an essential role in the proofs of the Feder--Vardi conjecture~\cite{BKOPP}.
At the same time, it covers  CSP-reformulations of many natural problems in qualitative reasoning, as well as all problems definable in MMSNP.
Every reduct of a countable finitely bounded homogeneous structure is uniquely described by an SNP sentence up to isomorphism, and its CSP is definable in monotone SNP.

Current approaches to the Bodirsky--Pinsker conjecture are problematic in that they mostly rely on the scarce pool of partial classification results for countable finitely bounded homogeneous structures; see, e.g.,~\cite{bodirsky2010complexity,bodirsky2015schaefer,kompatscher2016complexity,bodirsky2017complexity}.
Despite not having any direct consequences for the conjecture, our results provide evidence for the need for a fundamentally new language-independent approach to it.
Below we give two examples of obfuscation stemming from the ADP.

First, there is a folklore result from finite-domain CSP research stating that there exists a polynomial-time computable function mapping finite structures $\struct{B}_1$ over an arbitrary finite relational signature to finite structures $\struct{B}_2$ over a finite binary relational signature such that $\CSP(\struct{B}_1)$ and  $\CSP(\struct{B}_2)$ are polynomial-time equivalent~\cite{bulin2015finer,feder1998computational}.
In other words, the full complexity of finite-domain CSPs is already witnessed over binary signatures, and the reduction to binary signatures is efficient.
By Corollary~\ref{cor:hardness} and Proposition~\ref{cor:conexptime}, such a polynomial-time computable function is unlikely to exist between universal sentences representing finitely bounded homogeneous structures, unless it avoids the ADP. 

Second, there is the fact that, for every $\Phi$ as in Theorem~\ref{thm:inside_out}, the CSP of every structure whose age equals $\fm(\Phi)$ is trivial, independently on whether or not $\fm(\Phi)$ has the SAP.
The reason is that, if $\struct{B}$ is any structure whose age equals $\fm(\Phi)$ for $\Phi$ as in Theorem~\ref{thm:inside_out}, then $\Age(\struct{B})$ contains a structure $\struct{C}$ whose domain consists of a single element and where each relation contains one tuple;
every structure over the signature of $\Phi$ homomorphically maps to $\struct{C}$, and hence $\CSP(\struct{B})$ is trivial. 

Some initial steps towards a language-independent approach to the conjecture of Bodirsky and Pinsker were taken in the recent works of Mottet and Pinsker~\cite{mottet2024smooth} and Bodirsky and Bodor~\cite{bodirsky2021canonical}; however, they do not fully address the issues stemming from the ADP. 
We elaborate on this claim in Section~\ref{subs:csp_lore_deep}.

\subsection{Subtleties of the infinite-domain CSP}  \label{subs:csp_lore_deep}
In~2016, Bodirsky and Mottet~\cite{Bodirsky-Mottet} presented a first general link between the Feder--Vardi and the Bodirsky--Pinsker conjecture: they gave a general polynomial-time reduction from infinite-domain CSPs to finite-domain CSPs that unifies most of the tractability results in complexity classifications for infinite-domain CSPs
(the universal-algebraic proof of the complexity dichotomy for CSPs definable in MMSNP~\cite{bodirsky2021proof} being a prominent example).
The condition under which this general tool works is the following: the set of all canonical \emph{polymorphisms} of a structure $\struct{B}$ parametrizing a given CSP must be in a certain precise sense \emph{equationaly non-trivial}~\cite{mottet2024smooth,pinsker2022current}. 
Here, a polymorphism is simply a homomorphism from a finite power of a structure to itself.
However, there are several classification results for infinite-domain CSPs where this general reduction fails to explain all tractable cases, and where more refined approaches are necessary; we give two examples below.

The first example is the complexity dichotomy of Bodirsky and K\'ara~\cite{bodirsky2010complexity} for \emph{temporal CSPs}, i.e., for CSPs of structures with domain $\mathbb{Q}$ and whose relations are definable by a Boolean combination of formulas of the form $(x=y)$ or $(x<y)$.
These problems are very well understood; tractable temporal CSPs can always be solved by a divide-and-conquer algorithm which attempts to construct a (weak) linear order on the instance by repeatedly searching for a set of potential minimal elements among the input variables, where each instance of the search is performed using an oracle for a tractable finite-domain CSP. The latter is determined by the shape of the Boolean combinations. 
E.g., in the case of 
$$
\CSP(\mathbb{Q};\{(x,y,z)\in \mathbb{Q}^3 \mid (x=y<z)\vee (y=z<x)\vee (z=x<y)\}),
$$
solving the finite-domain CSP in question amounts to solving linear equations modulo $2$~\cite{bodirsky2010complexity,bodirsky2022descriptive}.
It is known that the tractability results from \cite{bodirsky2010complexity} cannot be obtained using the general polynomial-time reduction from~\cite{Bodirsky-Mottet}, because the set of all canonical polymophisms of $(\mathbb{Q};<)$ is \emph{equationally trivial}~\cite{mottet2024smooth}.

In~2022, Mottet and Pinsker introduced the machinery of \emph{smooth approximations}~\cite{mottet2022smooth,mottet2024smooth}, vastly generalizing the methods in~\cite{Bodirsky-Mottet}.
Intuitively, the idea behind the theory is that even if the set of canonical polymorphisms does not witness the polynomial-time tractability of the CSP, comparing it against the set of all polymorphisms might reveal some extra symmetry that can be utilized later.
The last section of their paper is devoted to temporal CSPs, and the authors manage to reprove a significant part of the dichotomy of Bodirsky and K\'{a}ra on just a few pages.
They achieve this by applying some of their general results to first-order expansions of  $(\mathbb{Q};<)$ and obtaining either \textup{\textsf{NP}}-hardness for the CSP, or one of the two types of symmetry that played a fundamental role in the original proof from~\cite{bodirsky2010complexity}.
The said symmetry can then be used to prove correctness of the divide-and-conquer reduction to a finite-domain CSP described above.
The issue is that this is essentially just a different interpretation of the algorithm of Bodirsky and K\'{a}ra~\cite{bodirsky2010complexity}, or rather its refinement by Bodirsky and Rydval~\cite{bodirsky2022descriptive}; it still depends on the ``base language'' $(\mathbb{Q};<)$ (see Proposition~3.1 in \cite{bodirsky2022descriptive} and the last section of~\cite{mottet2024smooth}). 
In contrast, the reduction from~\cite{Bodirsky-Mottet} only relies on algebraic properties of structures parametrizing CSPs under the assumption of finitely bounded homogeneity, and is therefore language-independent.
Mottet~\cite{mottet2025algebraic} recently announced uniform algorithms for temporal CSPs, based on a combination of \emph{sampling} and either \emph{local consistency} or \emph{singleton affine integer programming}.
This is a big leap from the algorithmic perspective;
though, the correctness proofs for said algorithms still use the ad-hoc framework of \emph{free sets} from~\cite{bodirsky2010complexity}.
%

A similar situation occurs in the case of phylogeny CSPs~\cite{bodirsky2017complexity} (our second example), which capture decision problems concerning the existence of a binary tree satisfying certain constraints imposed on its leaves.  
Tractable phylogeny CSPs are strikingly similar to tractable temporal CSPs; they can always be solved by a divide-and-conquer algorithm repeatedly searching for a subdivision of the input variables into two parts, which represent the two different branches below the root of a potential binary tree.
As with tractable temporal CSPs, each instance of the search is performed using an oracle for a tractable finite-domain CSP.
For tractable phylogeny CSPs, already the homogeneity of the base language is both sufficient and necessary for proving the correctness of the above described reduction to the finite-domain CSP (Theorem~6.13 and Lemma~6.12 in~\cite{bodirsky2017complexity}).
We can therefore speak of a case of extreme language-dependency.
Regarding the approach of Mottet~\cite{mottet2025algebraic} based on sampling, in contrast to linear orders, it is not clear how to efficiently sample over leaf structures: for every $n\in \mathbb{N}$, there are exponentially many non-isomorphic trees with $n$ leaves (but only a single linear order of length $n$).
 Temporal and phylogeny CSPs are special cases of CSPs of structures obtainable from the universal homogeneous binary tree~\cite{BBPP18} by specifying relations using first-order formulas.
 Achieving a complexity dichotomy in this context will require a non-trivial combination of~\cite{bodirsky2010complexity} and~\cite{bodirsky2017complexity}.

The language dependency issue of temporal and phylogeny CSPs reflects negatively on the attitude towards the Bodirsky--Pinsker conjecture.
It indicates that an optimal way of approaching the conjecture might
be by gaining a very good understanding of the class of reducts of finitely bounded homogeneous structures, e.g., through some sort of a classification.
It is unclear how realistic this prospect would be as model-theoretic properties often tend to be undecidable~\cite{cherlin2011forbidden,BraunfeldUndec} and this could include the amalgamation property.
The optimistic interpretation of our results points in the direction of decidability of the ADP, but with a very impractical lower bound. 
One possible way of avoiding the ADP when approaching the conjecture could be to simplify it and show that the full ADP is in fact not relevant.
An example of a direction in which such a simplification could go is given in~\cite{pinsker2025three}; where the scope of the Bodirsky--Pinsker conjecture is reducted to the CSPs of reducts of finitely bounded homogeneous structures \emph{without algebraicity}, which correspond (\cite[Theorem~4.3.5]{Book}) to reducts of finitely bounded strong amalgamation classes.
The restriction of the question in the ADP to whether $\fm(\Phi)$ has the SAP unfortunately retains the original complexity due to~Theorem~\ref{thm:SAP_lemma}.